\documentclass[%
 reprint,
 amsmath,amssymb,
 aps,
]{revtex4-2}

\usepackage{graphicx}
\usepackage{cancel}
\usepackage{dcolumn}
\usepackage{bm}
\usepackage{multirow} 
\usepackage{color}   
\usepackage[mathlines]{lineno}

\begin{document}

\preprint{APS/123-QED}

\title{Luminosity–Dependent Assembly Bias of Brightest Cluster Galaxies from Weak Lensing and Clustering}

\author{Zhenjie Liu}
  \email{liuzhj26@sjtu.edu.cn}
    \affiliation{School of Physics and Astronomy, Shanghai Jiao Tong University, Shanghai 200240, China}
\affiliation{Division of Physics and Astrophysical Science, Graduate School of Science, Nagoya University, Nagoya 464-8602, Japan}
\author{Hironao Miyatake}
 \email{hironao.miyatake@nagoya-u.jp}
\affiliation{Kobayashi-Maskawa Institute for the Origin of Particles and the Universe (KMI), Nagoya University, Nagoya, 464-8602, Japan}
\affiliation{Institute for Advanced Research, Nagoya University, Nagoya 464-8601, Japan}
\affiliation{Kavli Institute for the Physics and Mathematics of the Universe (WPI), University of Tokyo, 5-1-5, Kashiwanoha, 277-8583, Japan}

\author{Joop Schaye}
 \affiliation{Leiden Observatory, Leiden University, PO Box 9513, 2300 RA Leiden, the Netherlands}

\author{Matthieu Schaller}
\affiliation{Leiden Observatory, Leiden University, PO Box 9513, 2300 RA Leiden, the Netherlands}
\affiliation{Lorentz Institute for Theoretical Physics, Leiden University, PO box 9506, 2300 RA Leiden, the Netherland}

\author{Keitaro Ishikawa}
\affiliation{Division of Physics and Astrophysical Science, Graduate School of Science, Nagoya University, Nagoya 464-8602, Japan}

\author{Tomomi Sunayama}
\affiliation{Academia Sinica Institute of Astronomy and Astrophysics (ASIAA), No.1, Sec. 4, Roosevelt Rd, Taipei 106319, Taiwan, R.O.C}


\date{\today}

\begin{abstract}
Assembly bias, which is the variation in halo clustering at fixed mass driven by formation history, has long been predicted by numerical simulations but remains difficult to confirm observationally. Previous studies have reported evidence for halo assembly bias by dividing samples according to galaxy stellar mass using various methods. In this work, we present observational measurements of halo assembly bias based on the luminosity of spectroscopically confirmed brightest cluster galaxies (BCGs). Using cluster catalogs and shear measurements from the DESI Legacy Imaging Surveys, we employ a mass-dependent halo-bias model to disentangle halo bias from its underlying mass dependence in galaxy–galaxy lensing and clustering measurements. We confirm that brighter BCGs are less strongly clustered on large scales, with a relative bias ratio deviating from unity at the $\sim3\sigma$ level, suggesting the presence of assembly bias. Similar qualitative trends are also found in the FLAMINGO and MillenniumTNG hydrodynamical simulations, strengthening the connection between galaxy luminosity and halo formation history.
\end{abstract}

\maketitle


\section{Introduction}
In the standard cosmological model, the large-scale structure of the Universe is always traced by dark matter halos, which are the gravitational structures that host galaxies and clusters. The spatial distribution of halos is primarily determined by mass, with more massive halos typically exhibiting stronger clustering because they collapse from clustered peaks in the overdensity region of the initial density field. However, clustering behavior also depends on the formation history and environment, a phenomenon known as halo assembly bias (HAB; \cite{Sheth2004MNRAS, Gao2005MNRAS, Harker2006MNRAS, Wechsler2006ApJ, Dalal2008ApJ, Sunayama2016MNRAS, Lazeyras2017JCAP,Sunayama2022arXiv}). 
At high mass scales, early-forming halos exhibit weaker clustering, because they typically originate from high-curvature peaks that can exceed the density threshold even in the underdense environment. On the other hand, late-forming halos have stronger clustering, since they originate from low-curvature peaks and need help from the overdensity peaks to exceed the density threshold. These curvatures are imprinted as the halo concentration in the present Universe. At low mass scales, however, the trend becomes opposite, since low-mass halos neighboring high-mass halos, hence, in the overdense region, experience more tidal stripping, and then the concentration of such halos becomes higher. For detailed discussions of these trends, readers should refer to \citet{Dalal2008ApJ}. 

Furthermore, HAB affects not only halo clustering but also the formation and clustering of galaxies, leading to galaxy assembly bias (GAB)—the dependence of galaxy clustering on halo assembly history at fixed halo mass \citep{Chaves2016MNRAS}. For instance, red galaxies with low specific star formation rates are more likely to inhabit older halos than blue, star-forming galaxies, resulting in different clustering between the two populations \citep{Xu2020MNRAS}. Beyond galaxy clustering, GAB also manifests in the halo occupation distribution, as shown in studies such as \citet{Hearin2016MNRAS, Artale2018MNRAS, Bose2019MNRAS}. Ignoring assembly bias in modeling can lead to systematics in galaxy–halo connection studies and biased inferences of cosmological parameters \citep{Zentner2014MNRAS, McEwen2018MNRAS, McCarthy2019MNRAS, Paviot2024A&A, Yirong2025ApJ}.

In observations, some galaxy or galaxy cluster properties are commonly used as proxies for halo formation time to search for evidence of assembly bias (AB), which includes galaxy color \citep{Lin2016ApJ}, star formation activity \citep{Yang2006ApJ, Banerjee2023}, stellar mass / luminosity \citep{Wang2022MNRAS,Oyarz2024ApJ}, and average projected cluster radius \citep{Miyatake2016PhRvL, Dvornik2017MNRAS}, among others \citep{Niemiec2018MNRAS, Lin2022A&A, Sunayama2022arXiv, Zhang2025Natur}. However, many studies failed to detect significant AB signals, possibly due to the weak correlation between the chosen properties and the halo assembly history. While some works reported positive detections of AB \citep{Yang2006ApJ, Miyatake2016PhRvL}, these findings have been challenged due to potential systematic effects, such as halo mass estimation biases introduced by abundance matching \citep{Lin2016ApJ} or projection effects \citep{Zu2017MNRAS, Sunayama2019MNRAS,Sunayama2023MNRAS}. As a result, confirming the existence of AB still requires more robust observational methods and independent lines of evidence.

Several studies suggest that, at fixed halo mass, more massive central galaxies reside in earlier-forming halos \citep[e.g.,][]{Lim2016MNRAS, Matthee2017MNRAS, Xu2020MNRAS, Camargo2025MNRAS}, making central stellar mass a useful proxy for halo formation time. In observations, \citet{Zu2021MNRAS} split cluster samples of fixed abundance $\lambda$ into high- and low-stellar-mass subsamples, finding a $\sim 1.5 \sigma$ positive $M_{\ast}^{\rm BCG}$–$c$ correlation and a $\sim 1 \sigma$ evidence for halo assembly bias. In particular, they found that the low-$M_{\ast}^{\rm BCG}$, low-concentration clusters have $\sim 10\%$ higher large-scale bias than the high-$M_{\ast}^{\rm BCG}$, high-concentration ones at the same halo mass. \citet{Zu2022MNRAS} further proposed an analytic model to explain the observed phenomenon using the combination of halo assembly bias and central-satellite conformity. Using a similar sample-selection strategy, \citet{Sunayama2022arXiv} found comparable trends but also detected mild evidence for projection effects. In a complementary analysis, \citet{Oyarz2024ApJ} measured the number density around central galaxies of different stellar masses at fixed halo mass, providing additional evidence for halo assembly bias.

Motivated by previous studies, we construct luminosity-selected central galaxy samples without fixing halo mass in order to detect halo assembly bias, using a large sample of galaxy clusters and shear measurements from the Dark Energy Spectroscopic Instrument (DESI) Legacy Imaging Surveys. We measure the lensing and clustering of central galaxies with different luminosities, and correct for the halo bias induced by halo mass differences in order to isolate the assembly bias signal. Finally, we detect a significant assembly bias signal at the $\sim 3\sigma$ level, consistent with predictions from numerical simulations. This paper is organized as follows. In Section \ref{sec:data}, we describe the cluster catalog, shear catalog, and the simulation used for comparison. In Section \ref{sec:method}, we detail the modeling and measurement methods for excess surface density (ESD) and clustering. Section \ref{sec:results} presents the main observational results from lensing and clustering, and also includes an analysis of group-scale properties. In addition, we validate our findings using hydrodynamical simulations. Finally, we summarize our conclusions in Section \ref{sec:summary}.

\section{Data}\label{sec:data}
\subsection{Cluster Catalog}
\begin{figure}[ht!]
\centering
\includegraphics[width=0.35\textwidth]{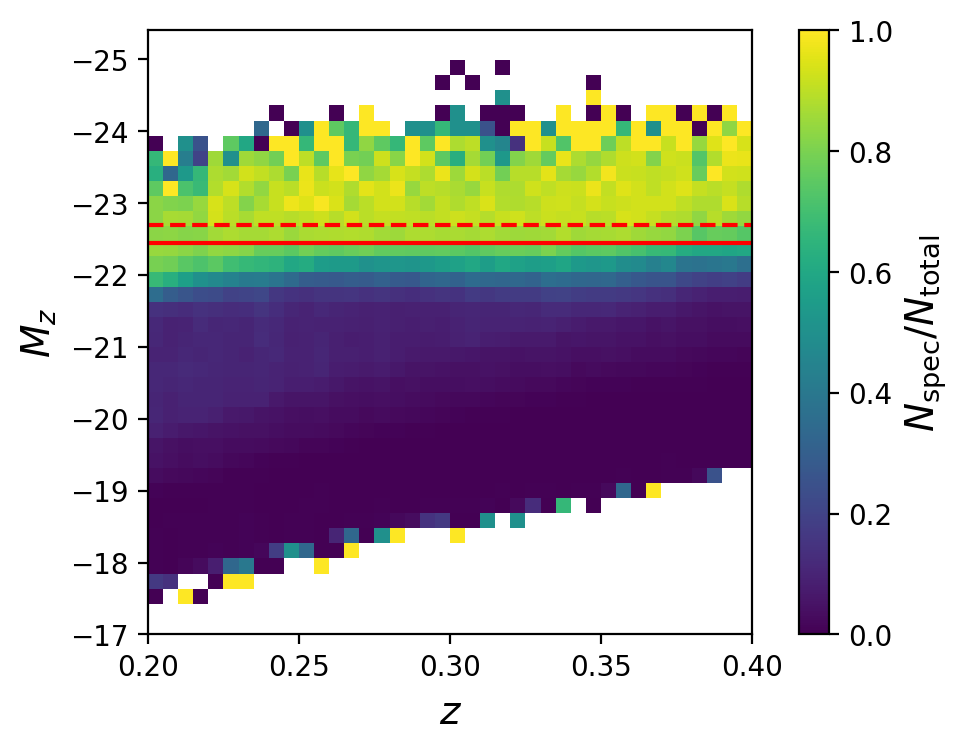}
\caption{Spectroscopic completeness of the BCG sample as a function of redshift $z$ and absolute $z$-band magnitude $M_z$. The color scale indicates the fraction of BCGs with spectroscopic redshifts, $N_{\mathrm{spec}}/N_{\mathrm{total}}$. The red solid and dashed lines mark the threshold at $-22.45$ and $-22.7$ respectively for two subsamples.} \label{fig:spec}
\end{figure}

The galaxy cluster catalog used in this study is based on the extended catalog developed by \citet{yang2021extended}, constructed from the DESI \citep{Dey2019AJ} Legacy Imaging Surveys Data Release 9. They identify galaxies belonging to the same dark matter halo based on spatial and redshift information, and use an iterative procedure that alternately updates group membership and halo mass via abundance matching, thereby ensuring a self-consistent and physically realistic cluster identification. The full catalog covers a total area of about 18,253 deg$^2$, but the shear catalog that we use covers a smaller area (see Sec.~\ref{shearcat}). To ensure overlap with the shear catalog and to remove irregular edges so as to mitigate the impact of survey geometry, and after removing exposures with a low number of stars and severe background contamination in shear catalog, the final observational sample spans an area of about 5,500 deg$^2$. Our analysis focuses on halo assembly bias as traced by the brightest cluster galaxies (BCGs), treating the BCGs as the central galaxies of the halos. We use galaxies in the redshift range $0.2<z<0.4$. To mitigate potential systematics in clustering measurements caused by photo-$z$ errors, we restrict our analysis to BCGs with spectroscopic redshifts. These redshifts are drawn from a comprehensive set of public surveys (e.g., BOSS, SDSS, GAMA), as detailed in Section 2.1 of \citet{yang2021extended}. To further address the issue of spectroscopic incompleteness, we follow the approach of \citet{zu2020arXiv} and \citet{Sunayama2022arXiv}, and present in Figure~\ref{fig:spec} the detection fraction of spectroscopic central galaxies as a function of $z$-band absolute magnitude $M_z$ and redshift. We find that for galaxies brighter than $M_z\lesssim -22.45$ (above the red solid line), the spectroscopic sample is nearly complete. Motivated by this, we divide the BCGs into two luminosity subsamples, $-22.7<M_z<-22.45$ and $M_z<-22.7$, so that the subsamples have a similar number of clusters. Table \ref{tab:sample} summarizes their sample sizes, comoving number density, and stellar masses.

\begin{table*} 
\centering  
\caption{Selection criteria for luminosity, sample sizes, comoving number densities and the mean stellar masses for the low-$L$ and high-$L$ BCGs subsamples in observation.} \label{tab:sample}
\begin{tabular}{l|c|ccc}
\hline  
 Samples& Luminosity cut& Number & $\bar{n}_g$/(Mpc/$h$)$^{-3}$ &${\rm log} \langle M_*/M_\odot \rangle$\\
[2.4pt]
\hline
low-$L$ & $-22.7<M_z<-22.45$ &25512 &$4.21 \times 10^{-5}$ &11.18\\
high-$L$&$M_z<-22.7$&23176&$3.83 \times 10^{-5}$  &11.35\\
 [2.4pt]
 \hline
\end{tabular}  
\end{table*} 

We use a random point sample that is at least ten times larger than the number of clusters to eliminate systematic errors and geometric effects in the lensing and correlation function measurements. Since the observed area of the clusters is irregular, we perform a regular trimming of the sky area to better match the random point catalog and the shear catalog, thereby reducing the systematic errors induced by geometric effects. We divide the sample into 100 subregions by using the K-means algorithm in Scikit-learn \citep{scikit-learn}, and apply the jackknife approach to obtain the covariance matrix for the measurements of both galaxy-galaxy lensing and correlation functions.

\subsection{Shear Catalog}\label{shearcat}
Our shear catalog is derived from the imaging data of the Dark Energy Camera Legacy Survey (DECaLS), processed using the Fourier\_Quad (FQ hereafter) shear measurement pipeline \citep{Zhang2022}. DECaLS is one of the three public projects in the DESI Legacy Imaging Surveys, designed to select targets for DESI under suboptimal seeing conditions ($\sim 1.5''$). This survey covers approximately 10,000 deg$^2$ of sky in the $g$, $r$, and $z$ bands. According to \citet{Zhang2022}, the $z$-band shear catalog exhibits the best quality, with the lowest multiplicative and additive biases. We therefore only use the $z$-band shear measurements for our analysis, which has a typical galaxy number density of approximately 3-5 per square arcmin. The photo-$z$ are obtained from \citet{Rongpu2021MNRAS}. To minimize the contamination from foreground galaxies due to photo-$z$ uncertainties, we restrict our selection to background galaxies with photo-$z$ values exceeding the lens redshift by 0.2, i.e., \(z_s > z_l + 0.2\). Additionally, galaxies with magnitudes greater than 21 are excluded, as their photo-$z$ tend to be less reliable due to their low brightness. We also discard shear measurements from the edges of the exposures where $\vert g_{f1,f2}\vert >0.0015$ (with $g_{f1}$ and $g_{f2}$ representing the two ellipticity components of the field distortion), from problematic chips, and from galaxies with a signal-to-noise ratio $\nu_F$ \citep{Li_2021} lower than 4. We test the shear measurement biases using the field-distortion method \citep{Zhang2019ApJ} and find that the multiplicative biases are consistent with zero at the 1$\sigma$ level.

\subsection{Hydrodynamics Simulations}
In this study, we employ the MillenniumTNG (MTNG; \cite{Hern2023MNRAS,Pakmor2023MNRAS}) and FLAMINGO \citep{Schaye2023MNRAS, Kugel2023MNRAS} hydrodynamical simulations to compare and validate our observational results. 

The MTNG project adopts the Planck 2015 cosmology \citep{Planck2016} and incorporates the IllustrisTNG galaxy formation model \citep{Springel2018MNRAS}, providing a powerful framework for investigating both large-scale structure and non-linear galaxy formation processes at high mass resolution. \citet{Bose2023MNRAS} show that MTNG can accurately reproduce the observed galaxy clustering as a function of stellar mass. The suite includes multiple hydrodynamical and N-body simulations with varying box sizes and resolution levels, some of which also incorporate a massive neutrino component.  
In this work, we use the largest full-physics run (MTNG-L500-4320-A), which features a comoving box size of $L=500 {\rm Mpc}/h \approx 738.12 \, {\rm Mpc}$ and is realized with $2 \times 4320^3$ resolution elements (dark matter particles and gas cells). This corresponds to a dark matter particle mass resolution of \(1.7 \times 10^8\,M_\odot\) and a baryonic cell mass resolution of \(3.1 \times 10^7\,M_\odot\). The stellar mass of a galaxy in MTNG is defined as the total mass of all star particles bound to its subhalo, and the halo mass is the total mass enclosed in a sphere whose mean density is 200 times the mean density of the universe. For our analysis, we use the snapshot at $z = 0.307$, which closely matches the mean redshift of our observational sample.

The FLAMINGO project represents the largest cosmological hydrodynamical simulation suite to date, designed to overcome the limitations of dark matter-only simulations in modeling baryonic effects on small scales, as well as the restricted volume and parameter coverage of earlier hydrodynamical runs. In addition, FLAMINGO explicitly accounts for both massive neutrinos and baryonic feedback. The stellar and AGN feedback models have been calibrated to reproduce the observed $z=0$ galaxy stellar mass function and the gas fractions of low-redshift clusters. Here we use the ``L1\_m8'' simulation, the highest-resolution run of the suite, with a box size of 1 Gpc, resolved with $3600^3$ baryonic particles. The initial mean baryonic particle mass is $1.3 \times 10^8 M_\odot$, while the mean CDM particle mass is $7.06 \times 10^8 M_\odot$. Following \citet{Schaye2023MNRAS}, we characterize BCGs using properties measured within a 50 kpc spherical aperture, including stellar mass and luminosity. We adopt the snapshot at $z=0.3$ for consistency with our observational dataset.

\section{Methodology}\label{sec:method}
In this work, we measured the excess surface density (ESD) of the sample using galaxy-galaxy lensing and also their projected auto-correlation functions. Here, we present our methodology, including the model for ESD and the measurements for both the ESD and the clustering. We assume a Planck 2018 cosmology \citep{Planck2020} in the model.

\subsection{Model for ESD}\label{sec:model}
For the ESD model, we take into account the effects of stellar mass of galaxies, halo profiles, off-centering effects, and the 2-halo term. 

For the effect of stellar mass, relative to halos, the mass of galaxies is concentrated within a very small radius, allowing it to be treated as a point source for its ESD as follows
\begin{equation}\label{eq:ESD_Ms}
\Delta  \Sigma_*(R)=\frac{M_*}{\pi R^2}.
\end{equation}
The stellar mass $M_*$ of BCGs is derived using the \textsc{Kcorrect} code \citep{Blanton2007AJ}, 
which estimates stellar masses through broadband SED fitting based on the stellar population synthesis models of \citet{Bruzual2003} 
with a initial mass function from \citet{Chabrier2003PASP}. The mean stellar mass for each subsample is listed in Table \ref{tab:sample}. We adopt the NFW model \citep{navarro1997universal} to describe an individual halo profile, whose density is expressed as
\begin{equation}\label{eq:rhor}
\rho (r)=\frac{\rho_0}{(r/r_s)(1+r/r_s)^2}
\end{equation}
with $\rho_0=\rho_m \Delta_{\rm vir}/(3I)$, where $I=[\ln (1+c)-c/(1+c)]/c^3$. $r_{\rm vir}$ is the virial radius, $r_s$ is the scale radius and the concentration $c$ of halos is defined by $c= r_{\rm vir}/r_s$. In this paper, we define the halo mass with an overdensity of \( \Delta_{\rm vir} = 200 \) times the mean matter density \( \rho_m \) and the halo mass is given by $M_{\rm 200m}=\Delta_{\rm vir}(4 \pi /3) \rho_m r^3_{\rm vir}$. The unit of $M_{\rm 200m}$ for both observations and simulations in this work is ${\rm M_\odot}/h$. We employ the analytical expression for the surface density \( \Delta \Sigma_{\rm NFW} \) as presented in \citet{yang2006}.

Although many studies suggest that the BCG is a good indicator of the halo's center compared to luminosity-weighted or density-weighted centers \citep[e.g.,][]{Luo2018ApJ, Wang2022}, their results show that only about 50\% of BCGs truly reside at the gravitational centers of their host halos. And for massive halos, many studies found that the central fraction of BCGs is only about 30\%-40\% \cite[e.g.][]{Hikage2013, Wang2014, Hoshino2015, Lange2017}. This implies that a significant fraction of BCGs are satellites misidentified as centrals, which can bias weak-lensing analyses that assume the BCG marks the true halo center, leading to underestimated halo masses and misestimated halo concentrations. We therefore include the off-centering effect in the 1-halo-term ESD model, consisting of centered and miscentered components,
\begin{equation}\label{sigma1h}
\Delta \Sigma_{\rm 1h} (R)=f_{\rm c}\Delta \Sigma_{\rm NFW}(R) +(1-f_{\rm c})\Delta \Sigma_{\rm off}\left(R \right) 
\end{equation}
where $f_{\rm c}$ is the fraction of centered halos. $\Sigma_{\rm off}$ refers to mean surface density of miscentered halos, which can be derived from
\begin{equation}\label{eq:sigmaoff}
\begin{aligned}
\Sigma_{\text {off }}\left(R \right)&=\frac{1}{2 \pi} \int_0^{\infty} {\rm d} R_{\rm off} P\left(R_{\rm off} \right)   \int_0^{2 \pi} {\rm d} \theta\\
 &\times \Sigma_{\rm NFW} \left(\sqrt{R^2+R_{\rm off}^2+2 R R_{\rm off} \cos \theta}\right).
\end{aligned}
\end{equation}
$P\left(R_{\rm off} \right)$ is the distribution of offset centers on radius and it is generally assumed to be a Rayleigh distribution \citep{offcen2007}
\begin{equation}\label{Prs}
P\left(R_{\rm off} \right)=\frac{R_{\rm off}}{\sigma_s^2} \exp \left(-\frac{R_{\rm off}^2}{2 \sigma_s^2}\right)
\end{equation}
$\sigma_s$ represents the scatter of off-centered positions. In this work, we treat it as a free parameter and use the halo radius $r_{\rm vir}$ as the unit.

Following \citet{van2013} and \citet{Wang2022}, the surface density of the 2-halo term can be expressed as,
\begin{equation}\label{sigma2h}
 \Sigma_{\rm 2h}(R)=2 \rho_m \int_R^\infty [1+ \xi_{\rm gm}^{\rm 2h}(r)]\frac{r {\rm d}r}{\sqrt{r^2- R^2}}
\end{equation}
where 
\begin{equation}\label{xi_2h}
\xi_{\rm gm}^{\rm 2h}(r)= A_b b_h (M_{\rm 200m}, z) \zeta (r,z) f_{\rm exc}(r) \xi_{\rm mm}(r).
\end{equation}
Here, we define $A_b b_h (M_{\rm 200m}, z)$ as the halo bias for this sample, where $b_h (M_{\rm 200m}, z)$ is the halo bias modeled by \citet{tinker2010large} and $A_b$ is a free parameter. If assembly bias is present, $A_b$ will deviate from 1. $\zeta (r,z)$ is the radial bias function obtained in \citet{van2013},
\begin{equation}
\zeta(r, z) = 
\begin{cases} 
\zeta_0(r, z) & \text{if } r \geq r_{\psi} \\ 
\zeta_0(r_{\psi}, z) & \text{if } r < r_{\psi}
\end{cases}
\end{equation}
where
\begin{equation}
\zeta_0(r, z) = \frac{[1 + 1.17\xi_{\text{mm}}(r, z)]^{1.49}}{[1 + 0.69\xi_{\text{mm}}(r, z)]^{2.09}}
\end{equation}
and $r_{\psi}$ is obtained by solving the following function
\begin{equation}\label{eq:zeta}
    \log_{10} \left[ \zeta_0 (r_\psi, z) \, \xi_{\text{mm}} (r_\psi, z) \right] = 0.9.
\end{equation}
$\xi_{\rm mm}(r)$ is calculated by COLOSSUS \citep{diemer2018colossus} using a power spectrum from Boltzmann solver CAMB \citep{camb}. $f_{\rm exc}(r)$ describes the halo exclusion effect, where it equals 0 when $r<r_{\rm vir}$ and equals 1 otherwise.

Combining all components in Eqs.\ref{eq:ESD_Ms}-\ref{eq:zeta}, our ESD model can eventually be summarized as
\begin{equation}\label{sigmatot}
\Delta \Sigma (R)=\Delta  \Sigma_*(R)+\Delta  \Sigma_{\rm 1h}(R| M_{\rm 200m}, c,\sigma_s, f_c)+\Delta  \Sigma_{\rm 2h}(R | b_h)
\end{equation}
To enhance the accuracy of our model, we divide the redshift range of 0.2 to 0.4 into six bins and perform an integration on redshift based on their distribution of each lens sample.

In total, our model comprises five free parameters, halo mass $M_{\rm 200m}$, concentration $c$, normalized halo bias $A_b$, the scatter of off-centering $\sigma_s$ and the proportion of centered halos $f_c$, which we fit using Markov Chain Monte Carlo techniques \citep{mcmc2000}. The flat priors of the parameters in the fitting process are listed in Table \ref{tab:prior}.

\begin{table} 
\centering  
\caption{Flat prior ranges adopted for all free parameters in the ESD model during MCMC fitting and their descriptions.} \label{tab:prior}
\begin{tabular}{c|c|l}
\hline  
 Parameter &Prior&Description \\
[2.4pt]
\hline
log$M_{\rm 200m}$&[12.5,14.5]&halo mass \\
$c$&[0,50]&halo concentration  \\
$\sigma_s$ & [0.01,1.2] & the scatter of off-centering \\
$f_c$& [0,1] &well-centering fraction \\
$A_b$  &[0,5]&halo bias normalized by $b_h (M_{\rm 200m}, z)$\\
\hline
\end{tabular}  
\end{table} 

\subsection{ESD Measurements}\label{sec:PDF}
Galaxy-galaxy lensing measures the correlation between the position of an object and the surrounding shear, which can reflect the ESD around the object. In general, the relationship between shear and ESD is given by
\begin{equation}\label{eq:ESD}
\Delta \Sigma (R)=\overline{\Sigma} (<R)-\Sigma(R)=\Sigma_c \gamma_t(R),
\end{equation}
where $\overline{\Sigma}(<r)$ refers to the average surface density within a radius $r$. $\Sigma_c$ is the comoving critical surface density, defined as 
\begin{equation}\label{sigmac}
    \Sigma_c=\frac{c^2}{4 \pi G } \frac{D_{\rm s}}{D_{\rm l} D_{\rm l s}(1+z_{\rm l})^2}
\end{equation} 
Here, $c$ is the speed of light, $G$ is the gravitational constant, and $D_{\rm s}, D_{\rm l}$, and $D_{\rm ls}$ are the angular diameter distances for the lens, source, and lens-source systems, respectively.

The shear estimators and statistical methods used in this work are uniquely defined by the FQ method, from which our shear catalog is derived. This method describes each galaxy image with a set of five quantities: $G_1$, $G_2$, $N$, $U$, and $V$, where $G_i$ are analogous to the ellipticity components $e_i$, $N$ is a normalization factor, and $U$ and $V$ are additional correction terms (see \citet{Zhang2015JCAP} for more details). Shear estimates within the FQ framework are obtained using the PDF-symmetrization method (PDF-SYM, \citep{Zhang2017}), which determines the shear by adjusting the assumed underlying value until the PDF of the estimator becomes maximally symmetric. For ESD measurements at a given radius $r$, we construct the PDF of the tangential shear estimator, $P(G_t)$, where $G_t$ is rotated from $G_1$ and $G_2$. Given a trial ESD signal $\widehat{\Delta \Sigma}$, we shift each $G_t$ value to define a new estimator,
\begin{equation}\label{Ghat}
\hat{G}_t = G_t - \frac{\widehat{\Delta \Sigma}}{\Sigma_c}\left(N \pm U_t\right),
\end{equation}
and obtain the corresponding distribution $P(\hat{G}_t)$. When the trial value $\widehat{\Delta \Sigma}$ matches the true signal, $P(\hat{G}_t)$ becomes symmetric, providing the optimal ESD estimate. Finally, to mitigate boundary effects and residual systematics, we subtract the signal measured around random points, with the number of randoms set to 10 times that of the lens galaxies.

\subsection{Projected Auto-correlation Function Measurements}
To avoid complications caused by redshift space distortions (RSD), we measure the projected correlation function and it is given by
\begin{equation}\label{eq:wp}
   w_p(R) = 2 \int_{0}^{\Pi_{\text{max}}} d\Pi \xi(R, \Pi).
\end{equation}
Here, $R$ and $\Pi$ are the separations perpendicular to and along the line of sight respectively. $\Pi_{\text{max}}$ is the maximum projection distance, which is primarily set to 100 Mpc/$h$ in our work, while we also test 50 Mpc/$h$ to check projection effects. For the integration over $\Pi$, we do the summation of $\xi (R, \Pi)$ with $\Delta \Pi = 1{\rm Mpc}/h$ from 0 to $\Pi_{\rm max}$. $\xi(R, \Pi)$ is the three-dimensional correlation function, which is estimated by the Landy and Szalay estimator \citep{LS1993},
\begin{equation}
    \xi (R, \Pi) = \frac{\langle {\rm DD} \rangle - 2\langle {\rm DR} \rangle + \langle {\rm RR} \rangle}{\langle {\rm RR} \rangle},
\end{equation}
where $\langle {\rm DD} \rangle$, $\langle {\rm DR} \rangle$ and $\langle {\rm RR}\rangle$ are the normalized numbers of data-data, data-random, and random-random pairs at given separations. To avoid the halo exclusion effect, we calculate the projected correlation function for projected radii between 3 and 40 ${\rm Mpc}/h$ in 8 logarithmically spaced bins.

\section{Results}\label{sec:results}
In this section, we first use the ESD measurements to infer the halo properties and halo bias of the two luminosity-defined subsamples. We then present the corresponding relative bias ratio and further validate it through their projected correlation functions. We do not perform a joint fit of ESD and clustering, as the absolute halo bias is highly sensitive to the assumed cosmology and would require a much more involved modeling framework. Finally, we compare our results with those from hydrodynamic simulations.

\subsection{Halo Properties from ESD}

\begin{figure*}[ht!]
\centering
\includegraphics[width=1\textwidth]{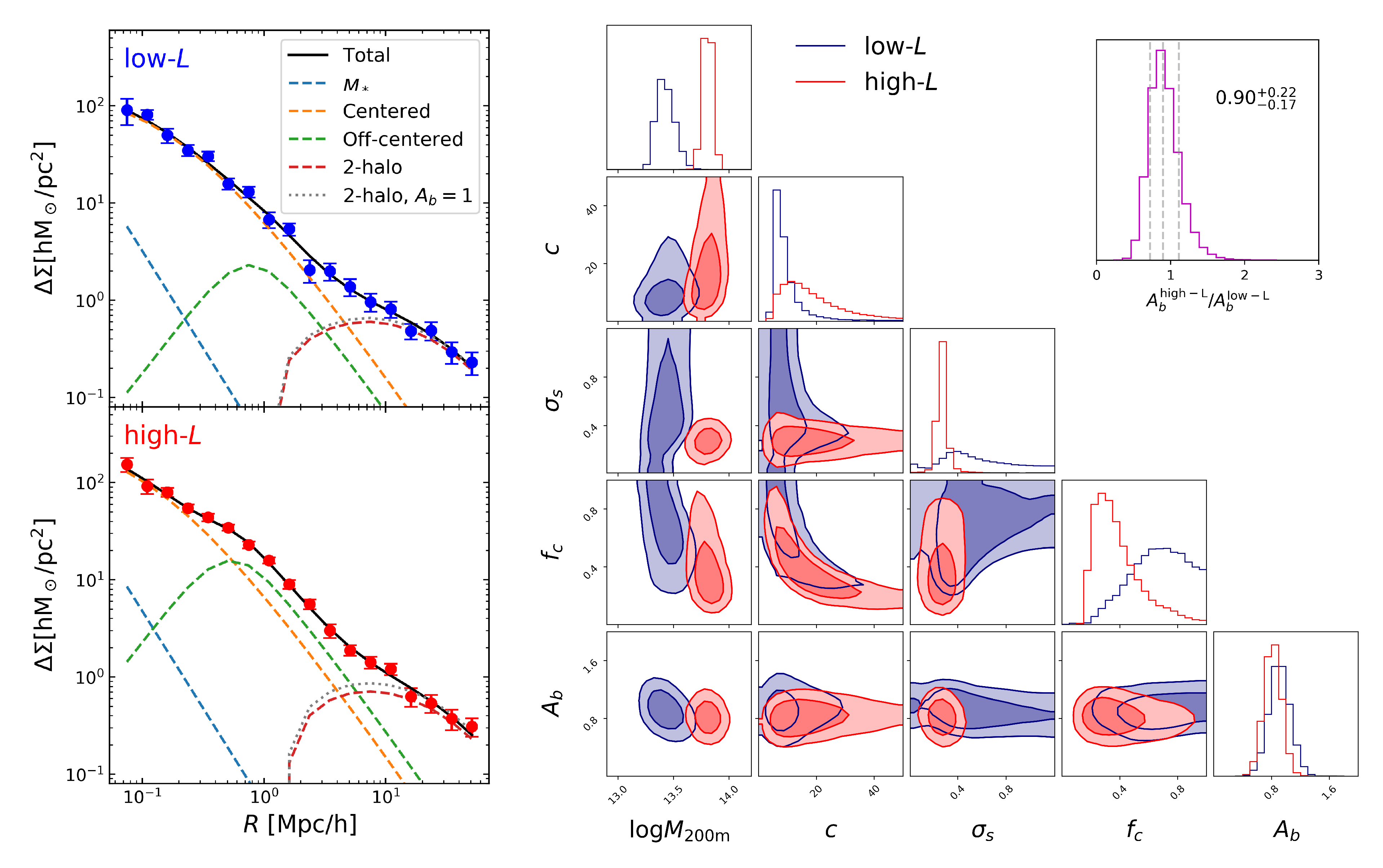}
\caption{Left panels: ESD profiles for the low-$L$ (upper panel, blue) and high-$L$ (lower panel, red) BCGs subsamples. The solid black lines show the best-fit models including all components, the colored dashed lines represent individual model components and the gray dotted lines show the 2-halo term with $A_b=1$ for comparison. Right panels: Posterior distributions of five free parameters and their 68\% and 95\% confidence contours in the ESD model for the two subsamples. The upper-right inset displays the posterior distribution of the ratio of normalized halo bias, $A_b^{{\rm high}-L}/A_b^{{\rm low}-L}$.
} \label{fig:ggl}
\end{figure*}

The left panel of Figure \ref{fig:ggl} shows the ESD measurements and best-fit models for the low-$L$ (upper panel, blue) and high-$L$ (lower panel, red) central galaxy subsamples. The dashed lines represent different components of the lensing model. The right panel of Figure \ref{fig:ggl} presents the posterior parameter constraints from the ESD fits, with numerical results listed in Table \ref{tab:gglresult}. Given the parameter degeneracies visible in Figure \ref{fig:ggl}, we check the robustness of our marginalized median estimates against prior volume or parameter space projection effects, by comparing the results from the maximum \textit{a posteriori} (MAP) listed in parentheses in Table \ref{tab:gglresult}. The close agreement between the median values and the MAP estimates confirms that our median results are robust against these effects.

As expected, the halos hosting brighter BCGs have higher masses. Typically, more massive halos are predicted to have lower concentrations since they undergo more accretion and major mergers. Interestingly, our results reveal that the high-$L$ sample exhibits a higher concentration, contrary to this expectation. This implies that the halos hosting brighter central galaxies may have formed earlier, and thus possess a higher concentration, consistent with the conclusion of \citet{Wang2022MNRAS}. Nevertheless, we note that the inferred concentration parameter $c$ is strongly degenerate with the off-centering parameter, as illustrated by the posterior contours. Therefore, while the higher $c$ for the high-$L$ sample is suggestive, its interpretation should be treated with caution and ideally corroborated by complementary or independent measurements.

The parameter $A_b$ (defined in Eq. \ref{xi_2h}) represents the halo bias normalized by the expected mass-dependent halo bias in the model, effectively removing the contribution from halo mass. Hence, deviations of $A_b$ from 1 can be regarded as evidence for assembly bias. We find that both subsamples exhibit $A_b$ values slightly below 1, with the high-$L$ sample showing an even lower value. The inset in the upper-right corner of Figure \ref{fig:ggl} displays the ratio of normalized halo bias between the two samples, $A_b^{\rm high-L}/A_b^{\rm low-L} = 0.90^{+0.22}_{-0.17}$, which is mildly below 1 but with modest statistical significance. The same lensing model has been shown to reproduce the expected halo bias for mass-selected DESI group samples \citep{Wang2022}, suggesting that the $A_b < 1$ values found here are unlikely to result from modeling deficiencies. Our result suggests that both subsamples may experience similar assembly bias, with a stronger effect for brighter central galaxies. 

Simulations predict that, at fixed mass, more concentrated halos tend to form earlier and host more massive central galaxies \citep{Matthee2017MNRAS,Xu2020MNRAS, Camargo2025MNRAS}. Moreover, \citet{Dalal2008ApJ} shows that massive halos originating from high-curvature peaks in the initial density field collapse earlier but are less strongly clustered than those associated with low-curvature peaks. Therefore, the mild trends we observe in both $c$ and $A_b$ are consistent with the expectations of assembly bias, indicating that more massive centrals and their host halos may have formed earlier.

\begin{table*} 
\centering  
\caption{Fitting results of ESD measurements for halo mass ${\rm log}M_{\rm 200m}$, concentration $c$, scatter in off-centering $\sigma_s$, the proportion of 
centered galaxies $f_c$ and halo bias $b_h$ for the low-$L$ and high-$L$ BCGs subsamples. The reported values correspond to the median and 68\% credible interval of the one-dimensional marginalized posterior distributions in Figure \ref{fig:ggl}. The MAP estimates are listed in parentheses, included as a check against parameter space projection effects. The $\chi^2/{\rm dof}$ value is evaluated at the median parameter values, with the corresponding value for the MAP point shown in parentheses.} \label{tab:gglresult}
\begin{tabular}{l|c|c|c|c|c|c}
\hline  
 Samples &${\rm log}M_{\rm 200m}$ & $c$ &$\sigma_s$ & $f_c$ & $A_b$ & $\chi^2/\nu$ \\
[2.4pt]
\hline
low-$L$ &$13.42^{+0.09}_{-0.07}$ (13.45) & 
$7.92^{+5.07}_{-1.96}$ (9.78) & 
$0.50^{+0.40}_{-0.23}$ (0.52) & 
$0.68^{+0.20}_{-0.21}$ (0.61) & 
$0.91^{+0.14}_{-0.14}$ (0.91) &  1.33 (1.32)\\
high-$L$&$13.81^{+0.04}_{-0.04}$ (13.81) & 
$16.11^{+12.70}_{-7.07}$ (16.93) & 
$0.27^{+0.04}_{-0.03}$ (0.26) & 
$0.35^{+0.20}_{-0.11}$ (0.34) & 
$0.82^{+0.12}_{-0.12}$ (0.84) &  0.61 (0.60)\\
 [2.4pt]
 \hline
\end{tabular}  
\end{table*} 

\subsection{Assembly bias from clustering}
\begin{figure}[ht!]
\centering
\includegraphics[width=0.4\textwidth]{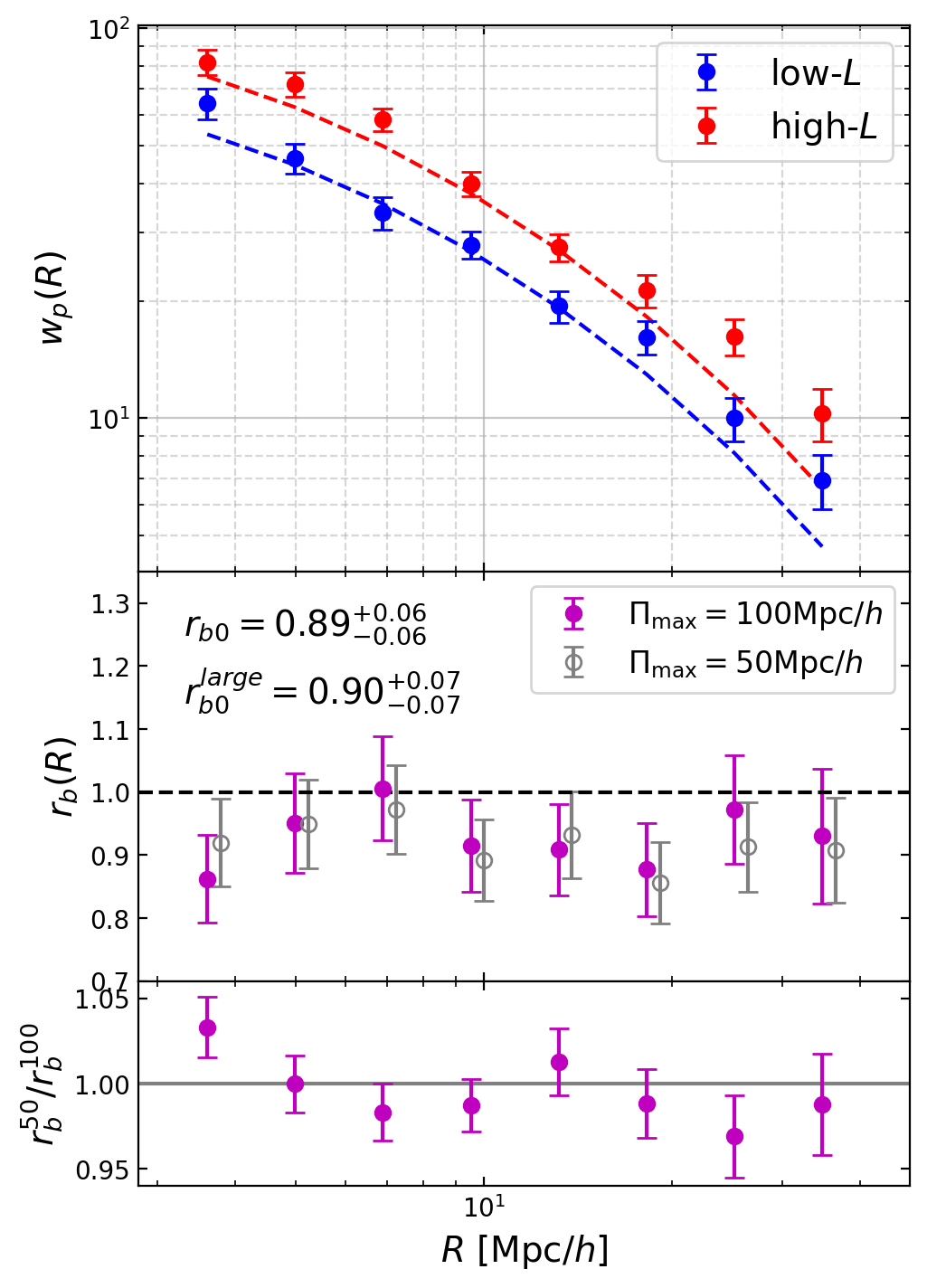}
\caption{Upper panel: Projected correlation functions $w_p(R)$ measured with $\Pi_{\max} = 50{\rm Mpc}/h$ for the low-$L$ (blue) and high-$L$ (red) BCG subsamples. The dashed curves show the corresponding theoretical predictions computed using the halo bias inferred from the ESD measurements. Middle panel: The relative bias ratio $r_b$ (defined in Eq.\ref{eq:rb}) between the two subsamples. A value of unity indicates no assembly bias. The error bars include the propagated uncertainties of halo masses measured via galaxy–galaxy lensing. Purple solid circles show the results using $\Pi_{\max} = 50 \, {\rm Mpc}/h$, while gray open circles represent the results with $\Pi_{\max} = 100 \, {\rm Mpc}/h$. The $R$ positions of the grey points are slightly shifted for visual clarity. $r_{b0}$ denotes the constant-fit value obtained from the purple data points, while $r_{b0}^{\rm large}$ corresponds to the same fit performed on those points with $R\gtrsim 10\,  {\rm Mpc/}h$. Lower panel: The ratio of $r_b(R)$ obtained with $\Pi_{\max} = 100\,{\rm Mpc}/h$ to that with $\Pi_{\max} = 50\,{\rm Mpc}/h$. The consistency with one demonstrates the robustness of our measurements against projection effects.} \label{fig:wp}
\end{figure}

Since the evidence for assembly bias in ESD is not significant, we further examine the projected auto-correlation functions of the two subsamples, as shown in Figure \ref{fig:wp}. The dashed curves show the theoretical predictions of $w_p(R)$ computed using the Planck 2018 cosmology, together with the halo masses and halo biases inferred from our ESD analysis. The slight mismatch between the data and the predictions may arise from the assumed cosmology. However, our analysis focuses on the relative bias ratio, for which the cosmology dependence is largely canceled. In Appendix \ref{app:regions}, we verify that the measured $w_p(R)$ is stable across different sky regions, indicating that our clustering results are not driven by spatial systematics.

In general, more massive halos are expected to be more strongly clustered, consistent with our finding that the high-$L$ sample exhibits a larger $w_p$ than the low-$L$ sample. To disentangle halo bias from its mass dependence, we compute the relative bias ratio,
\begin{equation}\label{eq:rb}
    r_b(R)=\frac{b_h^{{\rm low-}L}}{b_h^{{\rm high-}L}}\sqrt{\frac{w_p^{{\rm high-}L}}{w_p^{{\rm low-}L}}},
\end{equation}
where $b_h$ is the halo bias predicted by the model of \citet{tinker2010large} using the halo masses inferred from the ESD measurements. To test robustness, we also consider several alternative halo bias models and show (in Appendix \ref{app:bias_model}) that our results remain insensitive to the choice of bias prescription. 

The results, shown in the middle panel of Figure \ref{fig:wp}, indicate that $r_b(R)$ is systematically smaller than 1, implying that, after removing halo mass effects, brighter central galaxies are relatively less strongly clustered. The error bars already incorporate the propagated uncertainties of halo masses measured through galaxy–galaxy lensing. Since $r_b(R)$ does not display significant scale dependence, we model it with a constant bias ratio, obtaining $r_{b0} = 0.89 \pm 0.06$ from a least-squares fit. The result indicates evidence for assembly bias at the $\sim 3.2 \sigma$ level by $\chi^2$ statistic. To test the sensitivity of our result to the choice of scale range, we repeat the fit using only the clean 2-halo regime $R\gtrsim 10\,  {\rm Mpc/}h$. This yields $r_{b0}^{\rm large} = 0.90 \pm 0.07$, corresponding to a $\sim 2\sigma$ deviation from unity and in agreement with our baseline result. These results also align with the constraints inferred from the ESD analysis, although we note that the two measurements are not fully independent. Additionally, the observed trend is also consistent with the findings of \citet{Zu2021MNRAS, Zu2022MNRAS}, who detected similar signatures of assembly bias in more massive SDSS redMaPPer clusters.

To ensure our results are not contaminated by systematic errors, we test for the influence of projection effects, a key issue in optically selected cluster catalogs. This effect can cause an artificial, anisotropic boost in the large-scale clustering signal, potentially mimicking a physical signal like halo assembly bias. We follow the diagnostic method proposed by \citet{Sunayama2022arXiv} and \citet{Sunayama2023MNRAS}, which involves measuring the clustering signal using two different line-of-sight integration lengths ($\Pi_{\rm max}$) to assess this effect. As shown in the middle panel of Figure \ref{fig:wp}, we present the measurements for $\Pi_{\rm max}=100 \,{\rm Mpc}/h$ (purple) and $\Pi_{\rm max}=50 \,{\rm Mpc}/h$ (gray), which demonstrate excellent agreement. The ratio between the two, shown in the lower panel, remains consistent with unity across all scales, with a constant-fit value of $1.001\pm0.009$. This result provides strong evidence that our measurements are robust and not significantly biased by projection effects.

\subsection{Comparison with simulations}\label{sec:siml}
Cosmological simulations have established that, at fixed halo mass, central galaxy stellar mass correlates with halo formation redshift \citep[e.g.,][]{Matthee2017MNRAS, Artale2018MNRAS, Zehavi2018ApJ, Camargo2025MNRAS}. However, direct measurements of assembly bias related to stellar mass or luminosity remain scarce. Hydrodynamical simulations are usually too small for robust clustering statistics, while semi-analytic models on large-volume dark matter–only simulations cannot naturally capture assembly bias. Here, we leverage the large volumes of the MTNG and FLAMINGO hydrodynamical simulations to test for stellar mass–dependent assembly bias and compare with our observations.
\begin{table*}
\centering
\caption{Summary of selection criteria, sample sizes, comoving number densities, and average stellar and halo masses of the low-$L$ and high-$L$ BCG subsamples in the MTNG and FLAMINGO simulations. $r_{b0}$ is the constant fit to the relative bias ratio, indicating the strength of the assembly bias.}\label{tab:sample_siml}
\begin{tabular}{ccc|cccc|c}
\hline
 Simulations & Subsamples& Luminosity cut& Number & $\bar{n}_g$/(Mpc/$h$)$^{-3}$ &${\rm log} ( M_*/M_\odot)$&${\rm log} M_{\rm 200m}$ & $r_{b0}$\\
[2.4pt]
\hline
\multirow{2}{*}{MTNG} & low-$L$ & $-24.72<M_z<-25.15$ &5311 &$4.24 \times 10^{-5}$ &11.80&13.65& \multirow{2}{*}{$0.824 \pm 0.070$}\\
& high-$L$&$M_z<-25.15$&4779&$3.82 \times 10^{-5}$  &12.14&14.09& \\
 [2.4pt]
 \hline
 \multirow{2}{*}{FLAMINGO} & low-$L$ & $-24.31<M_z<-24.64$ &13461 &$4.26 \times 10^{-5}$ &11.71&13.58& \multirow{2}{*}{$0.849 \pm 0.067$}\\
 & high-$L$&$M_z<-24.64$&12098&$3.83 \times 10^{-5}$  &11.92&14.00& \\
 [2.4pt]
 \hline
\end{tabular}
\end{table*}

\begin{figure}[ht!]
\centering
\includegraphics[width=0.4\textwidth]{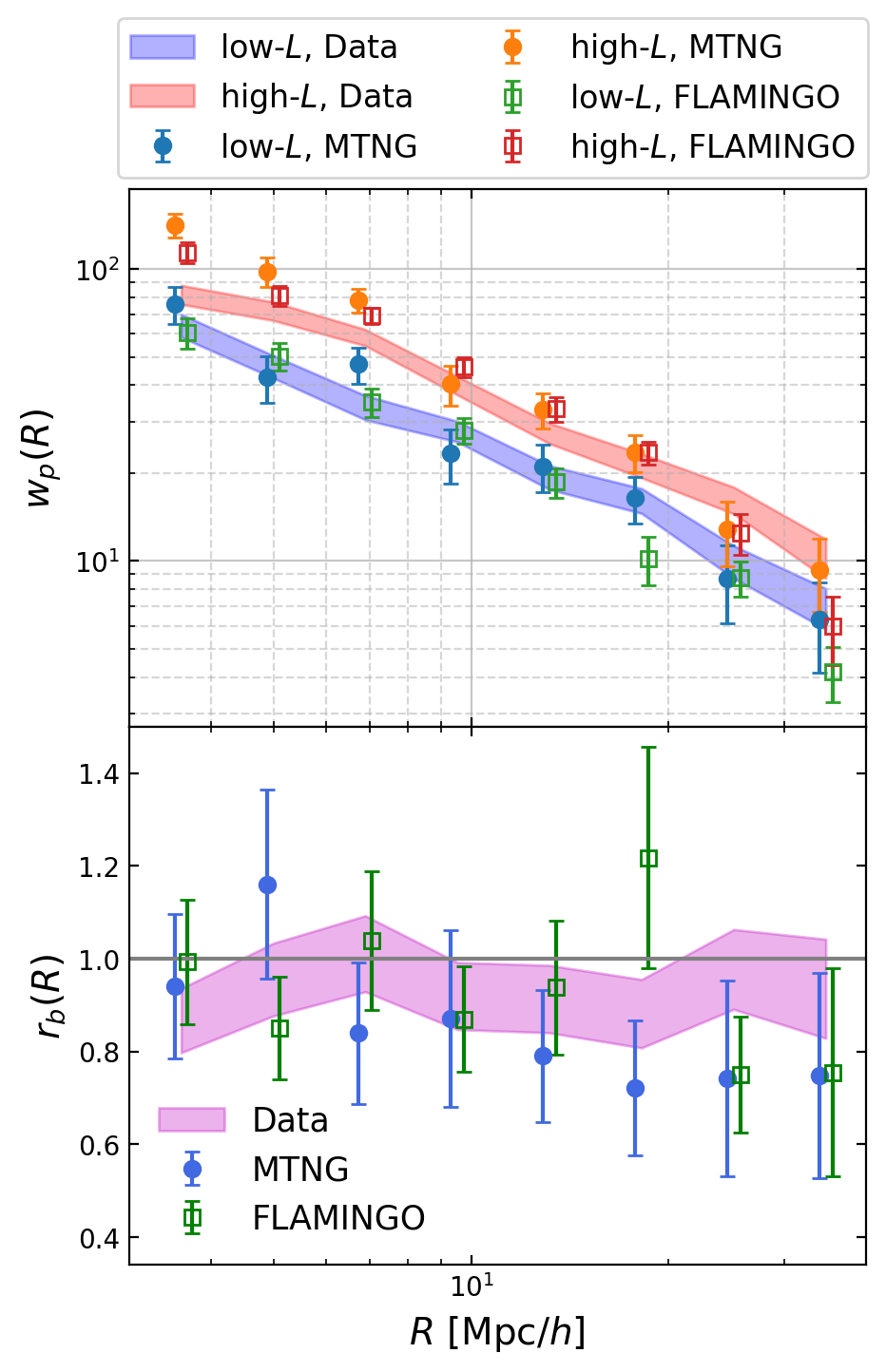}
\caption{The upper panel shows the projected correlation functions of the high-$L$ and low-$L$ BCG subsamples for MTNG and FLAMINGO hydrodynamical simulations; the lower panel shows the corresponding relative bias ratio $r_b(R)$ between the two subsamples. The shaded regions represent the corresponding observational results shown in Figure \ref{fig:wp}. A small horizontal offset is applied to the data points for clarity.
} \label{fig:wp_simu}
\end{figure}

Due to the different apertures used to measure luminosities in the simulations, the comparison with the observed absolute magnitudes is not straightforward. Instead, we adjust the luminosity cuts to match the comoving number densities of the observational samples, which are provided in Table \ref{tab:sample}, and the selection criteria for MTNG and FLAMINGO listed in Table~\ref{tab:sample_siml}. A detailed comparison of the observational and simulated samples is provided in Appendix \ref{app:sample}. To estimate the projected correlation functions, the simulation box is partitioned into $5 \times 5 \times 2$ subboxes, corresponding to $5 \times 5$ divisions on the projection plane and 2 divisions along the line of sight, and the jackknife resampling technique is applied. The correlation functions are computed in redshift space. The impact of RSD on the assembly bias measurements is also investigated, as discussed in Appendix~\ref{app:rsd}.

Figure~\ref{fig:wp_simu} shows the projected correlation functions $w_p(R)$ for the low-$L$ and high-$L$ subsamples in both simulations, as well as the relative bias ratio $r_b$ shown in the lower panel. For comparison, the shaded regions denote the corresponding observational results. In both MTNG and FLAMINGO, we find that $r_b(R)$ is systematically below one, as also indicated by the constant-fit results of the relative bias ratio in Table \ref{tab:sample_siml}. This suggests weaker clustering for the brighter central galaxies after removing the effect of halo mass. We assess the consistency between the observed bias ratio and the two hydrodynamical simulations by computing a $\chi^2$ statistic with the combined covariance. For MTNG, we find $\chi^2 /\nu=0.66$ with 8 degrees of freedom ($p=0.73$), demonstrating excellent agreement with our observations. The FLAMINGO prediction yields $\chi^2/\nu=1.02$ ($p=0.42$), which is still statistically consistent with the observations, although the deviation is somewhat larger compared to MTNG. We also show the constant fits of $r_b(R)$ for both simulations in Table \ref{tab:sample_siml}, which are systematically lower than unity and mutually consistent. These results qualitatively indicate similar assembly bias trends to those observed in the data.

\section{Summary and Conclusions}\label{sec:summary}
In this work, we investigate luminosity-dependent assembly bias of BCGs by combining weak lensing and clustering measurements. Using a large sample of galaxy clusters from the DESI Legacy Imaging Surveys with spectroscopically confirmed BCGs, we construct two subsamples based on luminosity. From the lensing measurements, we find that both BCG subsamples exhibit halo bias values slightly below the expected mass-dependent prediction, with the higher-luminosity subsample showing stronger hints of assembly bias, though with limited statistical significance. The clustering analysis provides clearer evidence: after correcting for halo mass effects and including their uncertainties, we obtain a mean relative bias ratio of $r_{b0}=0.89\pm0.06$, corresponding to a $3.2\sigma$ deviation from unity. These results agree with those inferred from the ESD analysis, demonstrating the robustness of our approach. 

The simulation analysis shows qualitatively similar trends. In both MTNG and FLAMINGO, we find $r_b$ systematically below unity, consistent in trend with the observational results and suggestive of a luminosity-dependent assembly bias. These similarities between the data and simulations point to a possible physical connection between BCG luminosity and halo formation history. A natural interpretation is that brighter central galaxies tend to reside in halos that formed earlier, and such early-forming halos are less strongly clustered at fixed mass, thereby leading to the lower halo bias we detect. 

Our results highlight the importance of considering galaxy assembly bias in interpreting halo–galaxy connections and cosmological observables. Future work with larger spectroscopic surveys and improved simulations will further refine these measurements and clarify the role of galaxy properties in tracing halo formation history.

\begin{acknowledgments}
We thank the MillenniumTNG collaboration for providing the simulation data used in this work. We thank Ying Zu and Surhud More for helpful suggestions on our work. Thank Mitsuyasu Yoshizaki for helping to download some of the data. Z.L. is supported by funding from China Scholarship Council. Z.L. and H.M are supported by JSPS KAKENHI Grant Numbers JP23H00108, JP23H04005, JP24KK0065. The computations in this paper were run on the $\pi$ 2.0 cluster supported by the Center of High Performance Computing at Shanghai Jiaotong University, and the ``Hubble'' computing cluster at Nagoya University. The FLAMINGO simulations used the DiRAC@Durham facility managed by the Institute for Computational Cosmology on behalf of the STFC DiRAC HPC Facility (\url{www.dirac.ac.uk}). The equipment was funded by BEIS capital funding via STFC capital grants ST/K00042X/1, ST/P002293/1, ST/R002371/1 and ST/S002502/1, Durham University and STFC operations grant ST/R000832/1. DiRAC is part of the National e-Infrastructure.

\end{acknowledgments}

\textbf{Data availability:} We made our results and data public on \url{https://github.com/load666/arXiv_2510.21377-results_data/tree/main}~\cite{Liu_2026_Data}.

\appendix

\section{The Impact of Halo Bias Models}
\label{app:bias_model}
\begin{figure*}[t]
    \centering
    \includegraphics[width=0.8\textwidth]{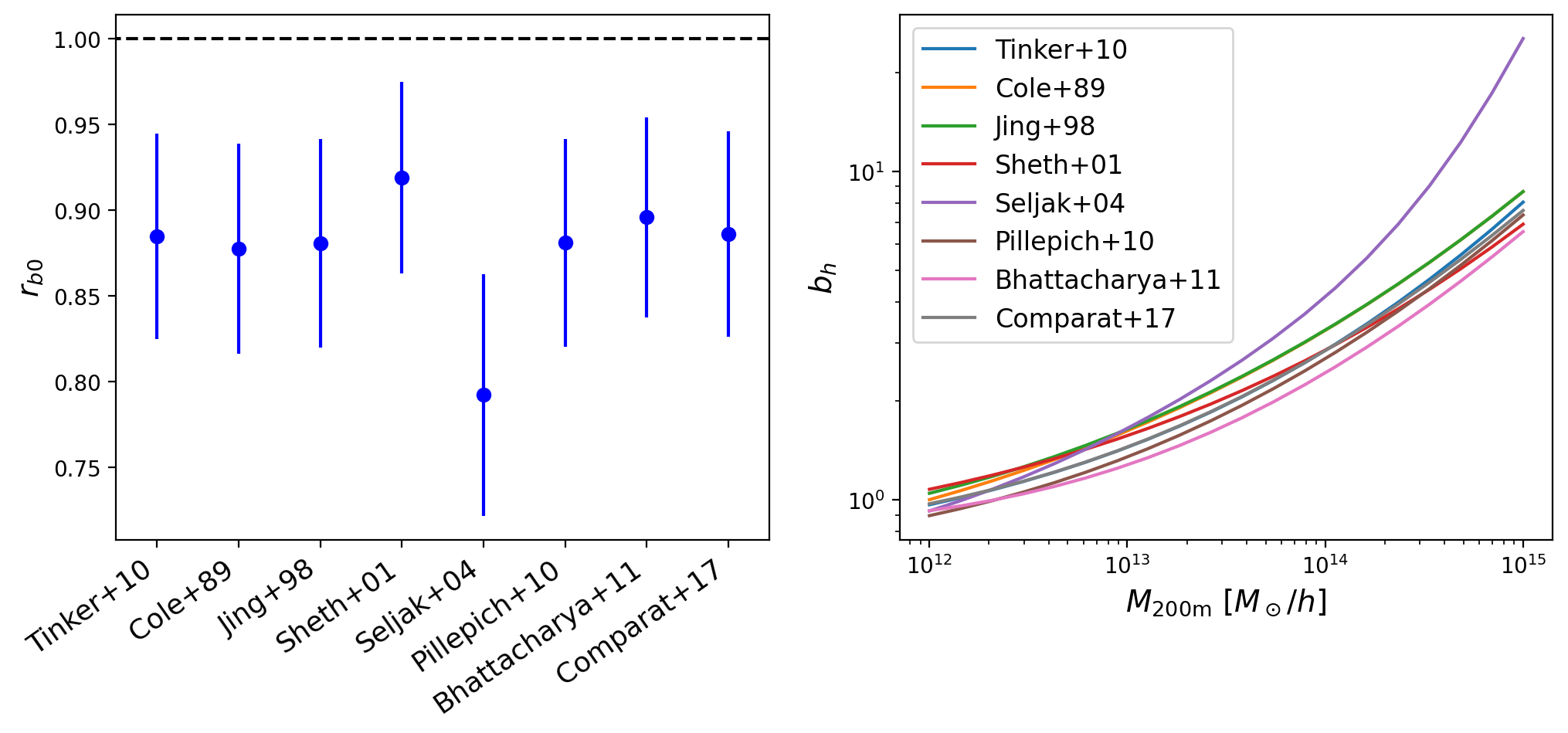}
    \caption{Left: Constant-fit results of the relative bias ratio, $r_{b0}$, obtained using different halo bias models.
    The horizontal axis labels the models in the order of \citet{tinker2010large}, 
\citet{Cole1989}, \citet{Jing1998}, \citet{Sheth2001}, \citet{Seljak2004}, \citet{Pillepich2010}, \citet{Bhattacharya2011}, 
and \citet{Comparat2017}. Right: Halo bias as a function of halo mass in different models.
    }
    \label{fig:bias_models}
\end{figure*}

In our analysis, when computing the relative bias ratio in Eq.\ref{eq:rb}, we need to adopt a halo bias model to remove the contribution of halo bias that is primarily driven by halo mass. Therefore, different halo bias prescriptions may potentially affect our measurement of assembly bias. To examine this, we evaluate $r_b(R)$ using several commonly adopted halo bias models available in the \texttt{Colossus} package \citep{Diemer2018ApJS}, including \citet{tinker2010large}, \citet{Cole1989}, \citet{Jing1998}, \citet{Sheth2001}, \citet{Seljak2004}, \citet{Pillepich2010}, \citet{Bhattacharya2011}, and \citet{Comparat2017}. As shown in the left panel of Figure~\ref{fig:bias_models}, the constant-fit values $r_{b0}$ are consistent among different halo bias prescriptions. The model of \citet{Seljak2004} yields a slightly lower $r_{b0}$, which likely arises from its higher halo bias amplitude at high-mass end and stronger mass dependence compared to others as shown in the right panel of Figure~\ref{fig:bias_models}. All other models produce nearly identical results, confirming that our detection of assembly bias is insensitive to the choice of the halo bias model.

\section{Clustering measurements in separate sky regions}\label{app:regions}
\begin{table} 
\centering  
\caption{Sample sizes, comoving number densities and the mean stellar masses for the low-$L$ and high-$L$ BCGs subsamples in northern and southern sky regions.} \label{tab:subregion}
\begin{tabular}{l|cccc}
\hline  
 Subsamples& Number & $\bar{n}_g$/(Mpc/$h$)$^{-3}$ &${\rm log} \langle M_*/M_\odot \rangle$\\
[2.4pt]
\hline
low-$L$, northern  &18112 &$4.38 \times 10^{-5}$ &11.176\\
low-$L$, southern  &7400 &$4.21 \times 10^{-5}$ &11.181\\
 [2.4pt]
 \hline
high-$L$, northern &16728&$4.04 \times 10^{-5}$  &11.347\\
high-$L$, southern &6448&$3.67 \times 10^{-5}$  &11.351\\
 [2.4pt]
 \hline
\end{tabular}  
\end{table} 

To assess the possible impact of observational systematics that vary across the survey footprint, we repeat our projected clustering measurements in two widely separated sky regions. These regions correspond to the northern and southern parts of the DESI Legacy Survey footprint, which is naturally split into two large sky regions by the Galactic Plane. The information of subsamples are listed in Table \ref{tab:subregion}. Such a split provides a consistency test, as any sky-dependent systematics might manifest as differences between the two regions.

\begin{figure}[t]
    \centering
    \includegraphics[width=0.45\textwidth]{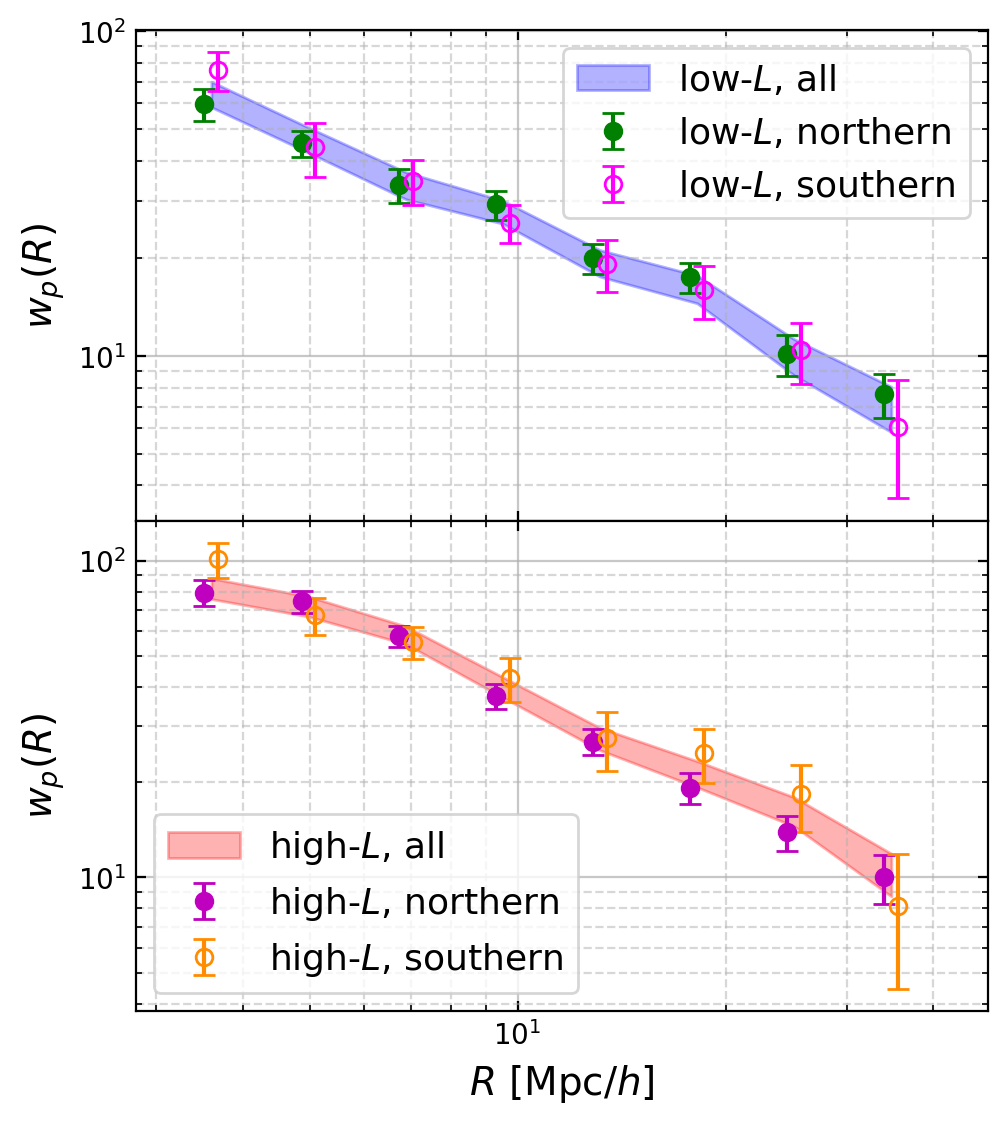}
    \caption{Projected correlation functions $w_p(R)$ for the low-$L$ (top) and high-$L$ (bottom) BCG subsamples from different sky regions. In each panel, the shaded band shows the measurement from the full samples, while the solid and hollow points correspond to the northern and southern sky regions separately. A small horizontal offset is applied to the data points for clarity.}
    \label{fig:subregion}
\end{figure}

Figure~\ref{fig:subregion} shows the projected correlation function $w_p(R)$ for the low-luminosity and high-luminosity samples, measured separately in northern and southern regions. For both luminosity bins, the two regions yield fully consistent clustering amplitudes and shapes within the statistical uncertainties. Moreover, the measurements in each region agree with the full-sample results. These tests demonstrate that the luminosity-dependent clustering measurements used in this work are not driven by sky-dependent observational systematics.

\section{Comparison of Observed and Simulated Luminosity–Halo Mass Relations}\label{app:sample}
\begin{figure*}[t]
    \centering
    \includegraphics[width=0.9\textwidth]{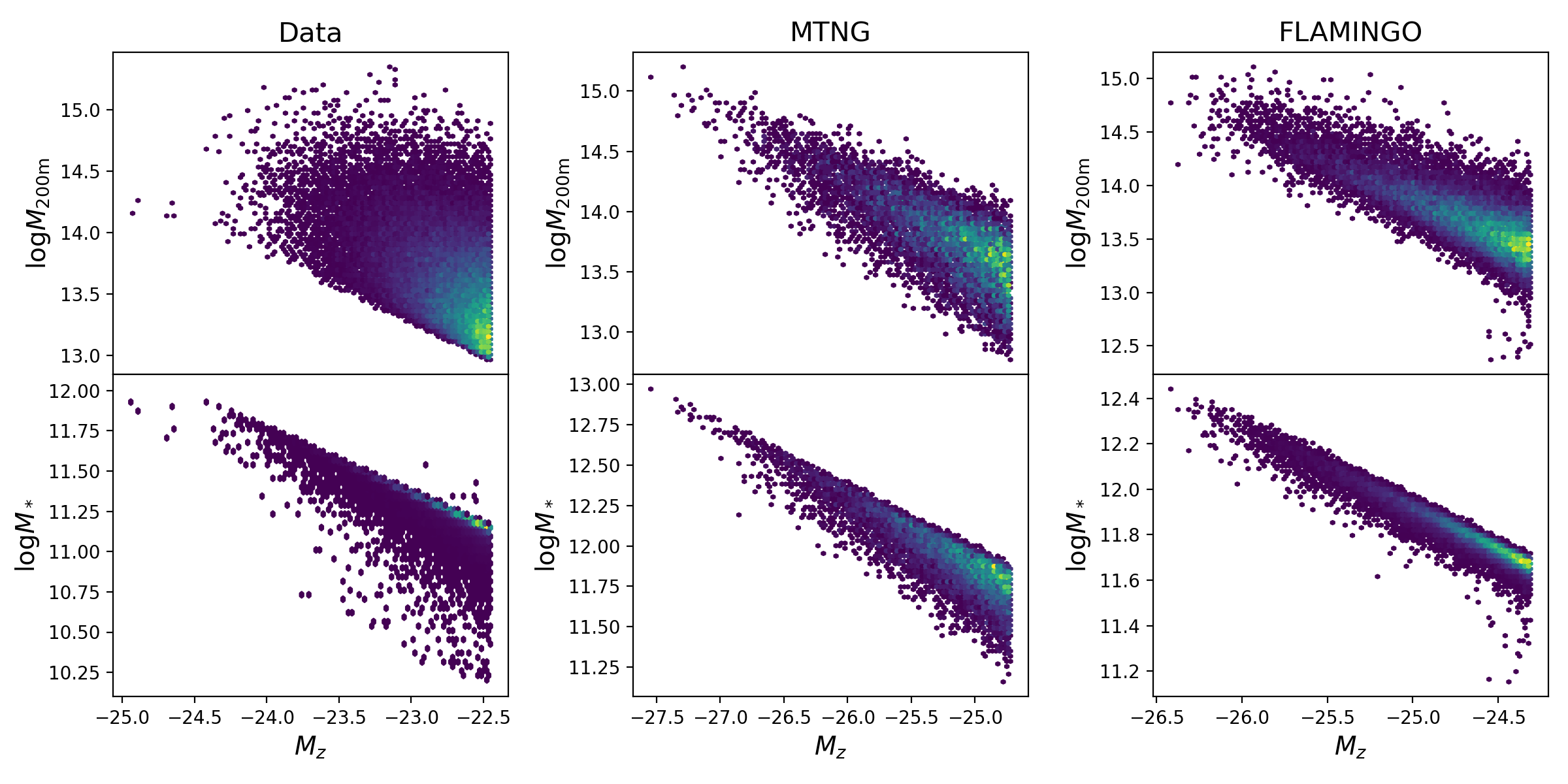}
    \caption{Distributions of halo mass (top row) and stellar mass (bottom row) as a function of z-band absolute magnitude $M_z$. The three columns show the observational sample (left), MTNG (middle), and FLAMINGO (right). For the data, halo masses are obtained by abundance matching from cluster catalog \citep{yang2021extended}.}
    \label{fig:Mh_mag}
\end{figure*}
To illustrate the differences between the observed BCG samples and their mock counterparts, Figure \ref{fig:Mh_mag} presents the distributions of halo mass and stellar mass as a function of the z-band absolute magnitude $M_z$ for the data, MTNG, and FLAMINGO. The halo masses for the observational sample in the left panel are obtained by abundance matching provided in the cluster catalog \citet{yang2021extended}. We note a broad scatter in the abundance-matched halo masses at fixed BCG luminosity in the data, whereas the hydrodynamical simulations exhibit much tighter luminosity–halo-mass relations. This large scatter in the observations may reflect satellite contamination or projection effects in the group catalog, where the identified BCG is not always the true central galaxy and it also manifests in our ESD analysis. By contrast, the central fractions in both MTNG and FLAMINGO are close to unity, naturally producing a much tighter correlation between luminosity and halo mass. Besides, we compute the median abundance-matched halo masses, ${\rm log}M_{\rm 200m}$, obtaining values of $13.40_{-0.24}^{+0.37}$ and $13.71_{-0.30}^{+0.42}$ for the low-$L$ and high-$L$ observational subsamples, respectively. These are in good agreement with our ESD-based mass estimates. This suggests that our halo masses from ESD are not significantly biased by these catalog-level systematics.

A second source of discrepancy arises from differences in how luminosity and stellar mass are defined across data and simulations. In MTNG, stellar masses are defined as the total mass of all star particles bound to a subhalo, while in FLAMINGO they are measured within a spherical aperture of 50 kpc. In the observations, luminosities are derived from photometric measurements and stellar masses are inferred from luminosity, making them sensitive to observational uncertainties and photometric systematics. These methodological differences lead to offsets in the stellar–halo mass relations between data and simulations.

Because of these differences in measurement definitions and modelling assumptions, achieving a one-to-one match between the full luminosity–halo mass distributions of the data and simulations would require a fully forward-modeled, survey-realistic mock catalog, which is beyond the scope of this work. Given our scientific goal and the current observational uncertainties, matching the number densities of the subsamples provides a robust and sufficient approach for constructing simulated analogs and testing the signatures of assembly bias.

\section{The Impact of Redshift-Space distortions}
\label{app:rsd}

\begin{figure}[t]
    \centering
    \includegraphics[width=0.48\textwidth]{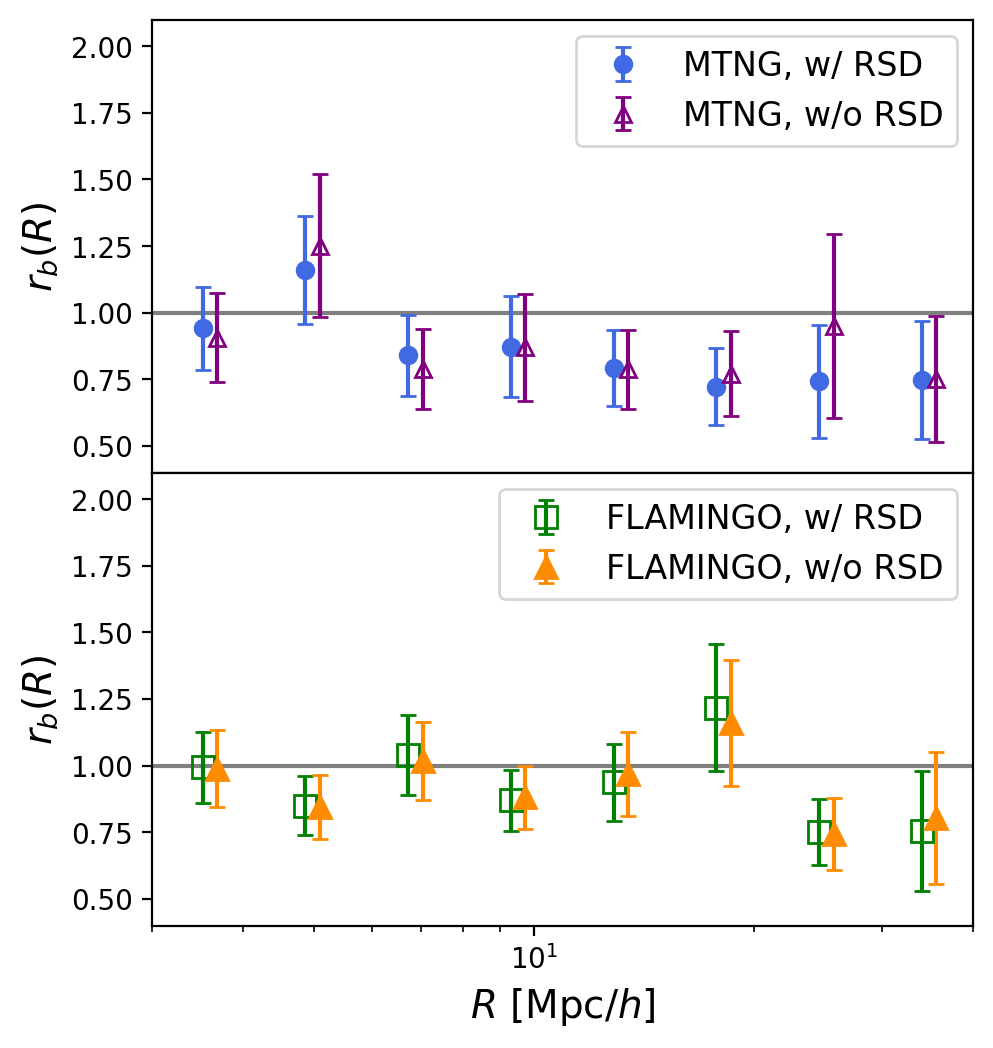}
    \caption{
    Comparison of the relative bias ratio $r_b(R)$ measured with and without including RSD in the MTNG and FLAMINGO simulations. A small horizontal offset is applied to the data points for clarity.
    }
    \label{fig:RSD_rb}
\end{figure}

In our measurement of the projected correlation function, we integrate the three-dimensional correlation function $\xi(R,\Pi)$ up to a line-of-sight separation of $\Pi_{\rm max}=100\,{\rm Mpc}/h$. 
Such a large integration length is expected to suppress most of the RSD effects, since peculiar velocities mainly alter clustering signals along the line of sight on smaller scales \citep{van2013MNRAS}. To explicitly verify this, we compare the relative bias ratio $r_b(R)$ both in real space (without RSD) and in redshift space (including RSD effects) for both MTNG and FLAMINGO simulations. As shown in Figure~\ref{fig:RSD_rb}, the results obtained with and without RSD are nearly identical in both simulations, with no systematic offset over the range of scales considered. This confirms that, given our projection length of $\Pi_{\rm max}=100\,{\rm Mpc}/h$, the effect of RSD on the relative bias ratio is negligible, and our conclusions regarding assembly bias are robust against the presence of RSD.

\bibliography{apssamp}

@ARTICLE{Banerjee2023,
       author = {{Banerjee}, Amrita and {Pandey}, Biswajit and {Nandi}, Anindita},
        title = "{Clustering and physical properties of the star-forming galaxies and AGN: does assembly bias have a role in AGN activity?}",
      journal = {arXiv e-prints},
         year = 2023,
        month = oct,
          eid = {arXiv:2310.12943},
        pages = {arXiv:2310.12943},
          doi = {10.48550/arXiv.2310.12943},
archivePrefix = {arXiv},
       eprint = {2310.12943},
 primaryClass = {astro-ph.GA},
       adsurl = {https://ui.adsabs.harvard.edu/abs/2023arXiv231012943B}
}

@ARTICLE{Zu2017MNRAS,
       author = {{Zu}, Ying and {Mandelbaum}, Rachel and {Simet}, Melanie and {Rozo}, Eduardo and {Rykoff}, Eli S.},
        title = "{On the level of cluster assembly bias in SDSS}",
      journal = {Monthly Notices of the Royal Astronomical Society},
         year = 2017,
        month = may,
       volume = {470},
       number = {1},
        pages = {551-560},
          doi = {10.1093/mnras/stx1264},
archivePrefix = {arXiv},
       eprint = {1611.00366},
 primaryClass = {astro-ph.CO},
       adsurl = {https://ui.adsabs.harvard.edu/abs/2017MNRAS.470..551Z}
}

@ARTICLE{Dvornik2017MNRAS,
       author = {{Dvornik}, Andrej and {Cacciato}, Marcello and {Kuijken}, Konrad and {Viola}, Massimo and {Hoekstra}, Henk and {Nakajima}, Reiko and {van Uitert}, Edo and {Brouwer}, Margot and {Choi}, Ami and {Erben}, Thomas and {Fenech Conti}, Ian and {Farrow}, Daniel J. and {Herbonnet}, Ricardo and {Heymans}, Catherine and {Hildebrandt}, Hendrik and {Hopkins}, Andrew M. and {McFarland}, John and {Norberg}, Peder and {Schneider}, Peter and {Sif{\'o}n}, Crist{\'o}bal and {Valentijn}, Edwin and {Wang}, Lingyu},
        title = "{A KiDS weak lensing analysis of assembly bias in GAMA galaxy groups}",
      journal = {Monthly Notices of the Royal Astronomical Society},
         year = 2017,
        month = jul,
       volume = {468},
       number = {3},
        pages = {3251-3265},
          doi = {10.1093/mnras/stx705},
archivePrefix = {arXiv},
       eprint = {1703.06657},
 primaryClass = {astro-ph.CO},
       adsurl = {https://ui.adsabs.harvard.edu/abs/2017MNRAS.468.3251D}
}

@ARTICLE{Dey2019AJ,
       author = {{Dey}, Arjun and {Schlegel}, David J. and {Lang}, Dustin and {Blum}, Robert and {Burleigh}, Kaylan and {Fan}, Xiaohui and {Findlay}, Joseph R. and {Finkbeiner}, Doug and {Herrera}, David and {Juneau}, St{\'e}phanie and {Landriau}, Martin and {Levi}, Michael and {McGreer}, Ian and {Meisner}, Aaron and {Myers}, Adam D. and {Moustakas}, John and {Nugent}, Peter and {Patej}, Anna and {Schlafly}, Edward F. and {Walker}, Alistair R. and {Valdes}, Francisco and {Weaver}, Benjamin A. and {Y{\`e}che}, Christophe and {Zou}, Hu and {Zhou}, Xu and {Abareshi}, Behzad and {Abbott}, T.~M.~C. and {Abolfathi}, Bela and {Aguilera}, C. and {Alam}, Shadab and {Allen}, Lori and {Alvarez}, A. and {Annis}, James and {Ansarinejad}, Behzad and {Aubert}, Marie and {Beechert}, Jacqueline and {Bell}, Eric F. and {BenZvi}, Segev Y. and {Beutler}, Florian and {Bielby}, Richard M. and {Bolton}, Adam S. and {Brice{\~n}o}, C{\'e}sar and {Buckley-Geer}, Elizabeth J. and {Butler}, Karen and {Calamida}, Annalisa and {Carlberg}, Raymond G. and {Carter}, Paul and {Casas}, Ricard and {Castander}, Francisco J. and {Choi}, Yumi and {Comparat}, Johan and {Cukanovaite}, Elena and {Delubac}, Timoth{\'e}e and {DeVries}, Kaitlin and {Dey}, Sharmila and {Dhungana}, Govinda and {Dickinson}, Mark and {Ding}, Zhejie and {Donaldson}, John B. and {Duan}, Yutong and {Duckworth}, Christopher J. and {Eftekharzadeh}, Sarah and {Eisenstein}, Daniel J. and {Etourneau}, Thomas and {Fagrelius}, Parker A. and {Farihi}, Jay and {Fitzpatrick}, Mike and {Font-Ribera}, Andreu and {Fulmer}, Leah and {G{\"a}nsicke}, Boris T. and {Gaztanaga}, Enrique and {George}, Koshy and {Gerdes}, David W. and {Gontcho}, Satya Gontcho A. and {Gorgoni}, Claudio and {Green}, Gregory and {Guy}, Julien and {Harmer}, Diane and {Hernandez}, M. and {Honscheid}, Klaus and {Huang}, Lijuan Wendy and {James}, David J. and {Jannuzi}, Buell T. and {Jiang}, Linhua and {Joyce}, Richard and {Karcher}, Armin and {Karkar}, Sonia and {Kehoe}, Robert and {Kneib}, Jean-Paul and {Kueter-Young}, Andrea and {Lan}, Ting-Wen and {Lauer}, Tod R. and {Le Guillou}, Laurent and {Le Van Suu}, Auguste and {Lee}, Jae Hyeon and {Lesser}, Michael and {Perreault Levasseur}, Laurence and {Li}, Ting S. and {Mann}, Justin L. and {Marshall}, Robert and {Mart{\'\i}nez-V{\'a}zquez}, C.~E. and {Martini}, Paul and {du Mas des Bourboux}, H{\'e}lion and {McManus}, Sean and {Meier}, Tobias Gabriel and {M{\'e}nard}, Brice and {Metcalfe}, Nigel and {Mu{\~n}oz-Guti{\'e}rrez}, Andrea and {Najita}, Joan and {Napier}, Kevin and {Narayan}, Gautham and {Newman}, Jeffrey A. and {Nie}, Jundan and {Nord}, Brian and {Norman}, Dara J. and {Olsen}, Knut A.~G. and {Paat}, Anthony and {Palanque-Delabrouille}, Nathalie and {Peng}, Xiyan and {Poppett}, Claire L. and {Poremba}, Megan R. and {Prakash}, Abhishek and {Rabinowitz}, David and {Raichoor}, Anand and {Rezaie}, Mehdi and {Robertson}, A.~N. and {Roe}, Natalie A. and {Ross}, Ashley J. and {Ross}, Nicholas P. and {Rudnick}, Gregory and {Safonova}, Sasha and {Saha}, Abhijit and {S{\'a}nchez}, F. Javier and {Savary}, Elodie and {Schweiker}, Heidi and {Scott}, Adam and {Seo}, Hee-Jong and {Shan}, Huanyuan and {Silva}, David R. and {Slepian}, Zachary and {Soto}, Christian and {Sprayberry}, David and {Staten}, Ryan and {Stillman}, Coley M. and {Stupak}, Robert J. and {Summers}, David L. and {Sien Tie}, Suk and {Tirado}, H. and {Vargas-Maga{\~n}a}, Mariana and {Vivas}, A. Katherina and {Wechsler}, Risa H. and {Williams}, Doug and {Yang}, Jinyi and {Yang}, Qian and {Yapici}, Tolga and {Zaritsky}, Dennis and {Zenteno}, A. and {Zhang}, Kai and {Zhang}, Tianmeng and {Zhou}, Rongpu and {Zhou}, Zhimin},
        title = "{Overview of the DESI Legacy Imaging Surveys}",
      journal = {The Astronomical Journal},
         year = 2019,
        month = may,
       volume = {157},
       number = {5},
          eid = {168},
        pages = {168},
          doi = {10.3847/1538-3881/ab089d},
archivePrefix = {arXiv},
       eprint = {1804.08657},
 primaryClass = {astro-ph.IM},
       adsurl = {https://ui.adsabs.harvard.edu/abs/2019AJ....157..168D}
}

@ARTICLE{Sunayama2023MNRAS,
       author = {{Sunayama}, Tomomi},
        title = "{Observational constraints of an anisotropic boost due to the projection effects using redMaPPer clusters}",
      journal = {Monthly Notices of the Royal Astronomical Society},
         year = 2023,
        month = jun,
       volume = {521},
       number = {4},
        pages = {5064-5076},
          doi = {10.1093/mnras/stad786},
archivePrefix = {arXiv},
       eprint = {2205.03233},
 primaryClass = {astro-ph.CO},
       adsurl = {https://ui.adsabs.harvard.edu/abs/2023MNRAS.521.5064S}
}

@ARTICLE{Zhang2019ApJ,
       author = {{Zhang}, Jun and {Dong}, Fuyu and {Li}, Hekun and {Li}, Xiangchong and {Li}, Yingke and {Liu}, Dezi and {Luo}, Wentao and {Fu}, Liping and {Li}, Guoliang and {Fan}, Zuhui},
        title = "{Testing Shear Recovery with Field Distortion}",
      journal = {The Astrophysical Journal},
         year = 2019,
        month = apr,
       volume = {875},
       number = {1},
          eid = {48},
        pages = {48},
          doi = {10.3847/1538-4357/ab1080},
archivePrefix = {arXiv},
       eprint = {1808.02593},
 primaryClass = {astro-ph.CO},
       adsurl = {https://ui.adsabs.harvard.edu/abs/2019ApJ...875...48Z}
}

@ARTICLE{Cole1989,
       author = {{Cole}, Shaun and {Kaiser}, Nick},
        title = "{Biased clustering in the cold dark matter cosmogony.}",
      journal = {Monthly Notices of the Royal Astronomical Society},
         year = 1989,
        month = apr,
       volume = {237},
        pages = {1127-1146},
          doi = {10.1093/mnras/237.4.1127},
       adsurl = {https://ui.adsabs.harvard.edu/abs/1989MNRAS.237.1127C}
}

@ARTICLE{Jing1998,
       author = {{Jing}, Y.~P.},
        title = "{Accurate Fitting Formula for the Two-Point Correlation Function of Dark Matter Halos}",
      journal = {The Astrophysical Journal Letters},
         year = 1998,
        month = aug,
       volume = {503},
       number = {1},
        pages = {L9-L13},
          doi = {10.1086/311530},
archivePrefix = {arXiv},
       eprint = {astro-ph/9805202},
 primaryClass = {astro-ph},
       adsurl = {https://ui.adsabs.harvard.edu/abs/1998ApJ...503L...9J}
}

@ARTICLE{Sheth2001,
       author = {{Sheth}, Ravi K. and {Mo}, H.~J. and {Tormen}, Giuseppe},
        title = "{Ellipsoidal collapse and an improved model for the number and spatial distribution of dark matter haloes}",
      journal = {Monthly Notices of the Royal Astronomical Society},
         year = 2001,
        month = may,
       volume = {323},
       number = {1},
        pages = {1-12},
          doi = {10.1046/j.1365-8711.2001.04006.x},
archivePrefix = {arXiv},
       eprint = {astro-ph/9907024},
 primaryClass = {astro-ph},
       adsurl = {https://ui.adsabs.harvard.edu/abs/2001MNRAS.323....1S}
}

@ARTICLE{Seljak2004,
       author = {{Seljak}, Uro{\v{s}} and {Warren}, Michael S.},
        title = "{Large-scale bias and stochasticity of haloes and dark matter}",
      journal = {Monthly Notices of the Royal Astronomical Society},
         year = 2004,
        month = nov,
       volume = {355},
       number = {1},
        pages = {129-136},
          doi = {10.1111/j.1365-2966.2004.08297.x},
archivePrefix = {arXiv},
       eprint = {astro-ph/0403698},
 primaryClass = {astro-ph},
       adsurl = {https://ui.adsabs.harvard.edu/abs/2004MNRAS.355..129S}
}

@ARTICLE{Pillepich2010,
       author = {{Pillepich}, Annalisa and {Porciani}, Cristiano and {Hahn}, Oliver},
        title = "{Halo mass function and scale-dependent bias from N-body simulations with non-Gaussian initial conditions}",
      journal = {Monthly Notices of the Royal Astronomical Society},
         year = 2010,
        month = feb,
       volume = {402},
       number = {1},
        pages = {191-206},
          doi = {10.1111/j.1365-2966.2009.15914.x},
archivePrefix = {arXiv},
       eprint = {0811.4176},
 primaryClass = {astro-ph},
       adsurl = {https://ui.adsabs.harvard.edu/abs/2010MNRAS.402..191P}
}

@ARTICLE{Bhattacharya2011,
       author = {{Bhattacharya}, Suman and {Heitmann}, Katrin and {White}, Martin and {Luki{\'c}}, Zarija and {Wagner}, Christian and {Habib}, Salman},
        title = "{Mass Function Predictions Beyond {\ensuremath{\Lambda}}CDM}",
      journal = {The Astrophysical Journal},
         year = 2011,
        month = may,
       volume = {732},
       number = {2},
          eid = {122},
        pages = {122},
          doi = {10.1088/0004-637X/732/2/122},
archivePrefix = {arXiv},
       eprint = {1005.2239},
 primaryClass = {astro-ph.CO},
       adsurl = {https://ui.adsabs.harvard.edu/abs/2011ApJ...732..122B}
}

@ARTICLE{Comparat2017,
       author = {{Comparat}, Johan and {Prada}, Francisco and {Yepes}, Gustavo and {Klypin}, Anatoly},
        title = "{Accurate mass and velocity functions of dark matter haloes}",
      journal = {Monthly Notices of the Royal Astronomical Society},
         year = 2017,
        month = aug,
       volume = {469},
       number = {4},
        pages = {4157-4174},
          doi = {10.1093/mnras/stx1183},
archivePrefix = {arXiv},
       eprint = {1702.01628},
 primaryClass = {astro-ph.CO},
       adsurl = {https://ui.adsabs.harvard.edu/abs/2017MNRAS.469.4157C}
}

@ARTICLE{Diemer2018ApJS,
       author = {{Diemer}, Benedikt},
        title = "{COLOSSUS: A Python Toolkit for Cosmology, Large-scale Structure, and Dark Matter Halos}",
      journal = {The Astrophysical Journal Supplement Series},
         year = 2018,
        month = dec,
       volume = {239},
       number = {2},
          eid = {35},
        pages = {35},
          doi = {10.3847/1538-4365/aaee8c},
archivePrefix = {arXiv},
       eprint = {1712.04512},
 primaryClass = {astro-ph.CO},
       adsurl = {https://ui.adsabs.harvard.edu/abs/2018ApJS..239...35D}
}

@ARTICLE{Bruzual2003,
       author = {{Bruzual}, G. and {Charlot}, S.},
        title = "{Stellar population synthesis at the resolution of 2003}",
      journal = {Monthly Notices of the Royal Astronomical Society},
         year = 2003,
        month = oct,
       volume = {344},
       number = {4},
        pages = {1000-1028},
          doi = {10.1046/j.1365-8711.2003.06897.x},
archivePrefix = {arXiv},
       eprint = {astro-ph/0309134},
 primaryClass = {astro-ph},
       adsurl = {https://ui.adsabs.harvard.edu/abs/2003MNRAS.344.1000B}
}

@ARTICLE{Chabrier2003PASP,
       author = {{Chabrier}, Gilles},
        title = "{Galactic Stellar and Substellar Initial Mass Function}",
      journal = {Publications of the Astronomical Society of the Pacific},
         year = 2003,
        month = jul,
       volume = {115},
       number = {809},
        pages = {763-795},
          doi = {10.1086/376392},
archivePrefix = {arXiv},
       eprint = {astro-ph/0304382},
 primaryClass = {astro-ph},
       adsurl = {https://ui.adsabs.harvard.edu/abs/2003PASP..115..763C}
}

@ARTICLE{camb,
       author = {{Lewis}, Antony and {Challinor}, Anthony and {Lasenby}, Anthony},
        title = "{Efficient Computation of Cosmic Microwave Background Anisotropies in Closed Friedmann-Robertson-Walker Models}",
      journal = {The Astrophysical Journal},
         year = 2000,
        month = aug,
       volume = {538},
       number = {2},
        pages = {473-476},
          doi = {10.1086/309179},
archivePrefix = {arXiv},
       eprint = {astro-ph/9911177},
 primaryClass = {astro-ph},
       adsurl = {https://ui.adsabs.harvard.edu/abs/2000ApJ...538..473L}
}

@ARTICLE{Matthee2017MNRAS,
       author = {{Matthee}, Jorryt and {Schaye}, Joop and {Crain}, Robert A. and {Schaller}, Matthieu and {Bower}, Richard and {Theuns}, Tom},
        title = "{The origin of scatter in the stellar mass-halo mass relation of central galaxies in the EAGLE simulation}",
      journal = {Monthly Notices of the Royal Astronomical Society},
         year = 2017,
        month = feb,
       volume = {465},
       number = {2},
        pages = {2381-2396},
          doi = {10.1093/mnras/stw2884},
archivePrefix = {arXiv},
       eprint = {1608.08218},
 primaryClass = {astro-ph.GA},
       adsurl = {https://ui.adsabs.harvard.edu/abs/2017MNRAS.465.2381M}
}

@ARTICLE{Bose2023MNRAS,
       author = {{Bose}, Sownak and {Hadzhiyska}, Boryana and {Barrera}, Monica and {Delgado}, Ana Maria and {Ferlito}, Fulvio and {Frenk}, Carlos and {Hern{\'a}ndez-Aguayo}, C{\'e}sar and {Hernquist}, Lars and {Kannan}, Rahul and {Pakmor}, R{\"u}diger and {Springel}, Volker and {White}, Simon D.~M.},
        title = "{The MillenniumTNG Project: the large-scale clustering of galaxies}",
      journal = {Monthly Notices of the Royal Astronomical Society},
         year = 2023,
        month = sep,
       volume = {524},
       number = {2},
        pages = {2579-2593},
          doi = {10.1093/mnras/stad1097},
archivePrefix = {arXiv},
       eprint = {2210.10065},
 primaryClass = {astro-ph.CO},
       adsurl = {https://ui.adsabs.harvard.edu/abs/2023MNRAS.524.2579B}
}

@ARTICLE{Blanton2007AJ,
       author = {{Blanton}, Michael R. and {Roweis}, Sam},
        title = "{K-Corrections and Filter Transformations in the Ultraviolet, Optical, and Near-Infrared}",
      journal = {The Astronomical Journal},
         year = 2007,
        month = feb,
       volume = {133},
       number = {2},
        pages = {734-754},
          doi = {10.1086/510127},
archivePrefix = {arXiv},
       eprint = {astro-ph/0606170},
 primaryClass = {astro-ph},
       adsurl = {https://ui.adsabs.harvard.edu/abs/2007AJ....133..734B}
}

@ARTICLE{Planck2020,
       author = {{Planck Collaboration} and {Aghanim}, N. and {Akrami}, Y. and {Ashdown}, M. and {Aumont}, J. and {Baccigalupi}, C. and {Ballardini}, M. and {Banday}, A.~J. and {Barreiro}, R.~B. and {Bartolo}, N. and {Basak}, S. and {Battye}, R. and {Benabed}, K. and {Bernard}, J.-P. and {Bersanelli}, M. and {Bielewicz}, P. and {Bock}, J.~J. and {Bond}, J.~R. and {Borrill}, J. and {Bouchet}, F.~R. and {Boulanger}, F. and {Bucher}, M. and {Burigana}, C. and {Butler}, R.~C. and {Calabrese}, E. and {Cardoso}, J.-F. and {Carron}, J. and {Challinor}, A. and {Chiang}, H.~C. and {Chluba}, J. and {Colombo}, L.~P.~L. and {Combet}, C. and {Contreras}, D. and {Crill}, B.~P. and {Cuttaia}, F. and {de Bernardis}, P. and {de Zotti}, G. and {Delabrouille}, J. and {Delouis}, J.-M. and {Di Valentino}, E. and {Diego}, J.~M. and {Dor{\'e}}, O. and {Douspis}, M. and {Ducout}, A. and {Dupac}, X. and {Dusini}, S. and {Efstathiou}, G. and {Elsner}, F. and {En{\ss}lin}, T.~A. and {Eriksen}, H.~K. and {Fantaye}, Y. and {Farhang}, M. and {Fergusson}, J. and {Fernandez-Cobos}, R. and {Finelli}, F. and {Forastieri}, F. and {Frailis}, M. and {Fraisse}, A.~A. and {Franceschi}, E. and {Frolov}, A. and {Galeotta}, S. and {Galli}, S. and {Ganga}, K. and {G{\'e}nova-Santos}, R.~T. and {Gerbino}, M. and {Ghosh}, T. and {Gonz{\'a}lez-Nuevo}, J. and {G{\'o}rski}, K.~M. and {Gratton}, S. and {Gruppuso}, A. and {Gudmundsson}, J.~E. and {Hamann}, J. and {Handley}, W. and {Hansen}, F.~K. and {Herranz}, D. and {Hildebrandt}, S.~R. and {Hivon}, E. and {Huang}, Z. and {Jaffe}, A.~H. and {Jones}, W.~C. and {Karakci}, A. and {Keih{\"a}nen}, E. and {Keskitalo}, R. and {Kiiveri}, K. and {Kim}, J. and {Kisner}, T.~S. and {Knox}, L. and {Krachmalnicoff}, N. and {Kunz}, M. and {Kurki-Suonio}, H. and {Lagache}, G. and {Lamarre}, J.-M. and {Lasenby}, A. and {Lattanzi}, M. and {Lawrence}, C.~R. and {Le Jeune}, M. and {Lemos}, P. and {Lesgourgues}, J. and {Levrier}, F. and {Lewis}, A. and {Liguori}, M. and {Lilje}, P.~B. and {Lilley}, M. and {Lindholm}, V. and {L{\'o}pez-Caniego}, M. and {Lubin}, P.~M. and {Ma}, Y.-Z. and {Mac{\'\i}as-P{\'e}rez}, J.~F. and {Maggio}, G. and {Maino}, D. and {Mandolesi}, N. and {Mangilli}, A. and {Marcos-Caballero}, A. and {Maris}, M. and {Martin}, P.~G. and {Martinelli}, M. and {Mart{\'\i}nez-Gonz{\'a}lez}, E. and {Matarrese}, S. and {Mauri}, N. and {McEwen}, J.~D. and {Meinhold}, P.~R. and {Melchiorri}, A. and {Mennella}, A. and {Migliaccio}, M. and {Millea}, M. and {Mitra}, S. and {Miville-Desch{\^e}nes}, M.-A. and {Molinari}, D. and {Montier}, L. and {Morgante}, G. and {Moss}, A. and {Natoli}, P. and {N{\o}rgaard-Nielsen}, H.~U. and {Pagano}, L. and {Paoletti}, D. and {Partridge}, B. and {Patanchon}, G. and {Peiris}, H.~V. and {Perrotta}, F. and {Pettorino}, V. and {Piacentini}, F. and {Polastri}, L. and {Polenta}, G. and {Puget}, J.-L. and {Rachen}, J.~P. and {Reinecke}, M. and {Remazeilles}, M. and {Renzi}, A. and {Rocha}, G. and {Rosset}, C. and {Roudier}, G. and {Rubi{\~n}o-Mart{\'\i}n}, J.~A. and {Ruiz-Granados}, B. and {Salvati}, L. and {Sandri}, M. and {Savelainen}, M. and {Scott}, D. and {Shellard}, E.~P.~S. and {Sirignano}, C. and {Sirri}, G. and {Spencer}, L.~D. and {Sunyaev}, R. and {Suur-Uski}, A.-S. and {Tauber}, J.~A. and {Tavagnacco}, D. and {Tenti}, M. and {Toffolatti}, L. and {Tomasi}, M. and {Trombetti}, T. and {Valenziano}, L. and {Valiviita}, J. and {Van Tent}, B. and {Vibert}, L. and {Vielva}, P. and {Villa}, F. and {Vittorio}, N. and {Wandelt}, B.~D. and {Wehus}, I.~K. and {White}, M. and {White}, S.~D.~M. and {Zacchei}, A. and {Zonca}, A.},
        title = "{Planck 2018 results. VI. Cosmological parameters}",
      journal = {Astronomy \& Astrophysics},
         year = 2020,
        month = sep,
       volume = {641},
          eid = {A6},
        pages = {A6},
          doi = {10.1051/0004-6361/201833910},
archivePrefix = {arXiv},
       eprint = {1807.06209},
 primaryClass = {astro-ph.CO},
       adsurl = {https://ui.adsabs.harvard.edu/abs/2020A&A...641A...6P}
}

@misc{Liu_2026_Data,
  author       = {Liu, Zhenjie},
  year         = {2026},
  title        = {{Luminosity--Dependent Assembly Bias of Brightest Cluster Galaxies from Weak Lensing and Clustering}},
  howpublished = {Zenodo},
  version      = {1.0.0},
  doi          = {10.5281/zenodo.18140021},
  note         = {v1.0.0, \url{https://doi.org/10.5281/zenodo.18140021}}
}

@ARTICLE{Yirong2025ApJ,
       author = {{Wang}, Yirong and {Zhai}, Zhongxu and {Yang}, Xiaohu and {Tinker}, Jeremy L.},
        title = "{Exploring the Signature of Assembly Bias and Modified Gravity Using Small-scale Clusterings of Galaxies}",
      journal = {\apj},
         year = 2025,
        month = nov,
       volume = {994},
       number = {1},
          eid = {51},
        pages = {51},
          doi = {10.3847/1538-4357/ae0c10},
archivePrefix = {arXiv},
       eprint = {2506.22737},
 primaryClass = {astro-ph.CO},
       adsurl = {https://ui.adsabs.harvard.edu/abs/2025ApJ...994...51W}
}

@ARTICLE{Chaves2016MNRAS,
       author = {{Chaves-Montero}, Jon{\'a}s and {Angulo}, Raul E. and {Schaye}, Joop and {Schaller}, Matthieu and {Crain}, Robert A. and {Furlong}, Michelle and {Theuns}, Tom},
        title = "{Subhalo abundance matching and assembly bias in the EAGLE simulation}",
      journal = {Monthly Notices of the Royal Astronomical Society},
         year = 2016,
        month = aug,
       volume = {460},
       number = {3},
        pages = {3100-3118},
          doi = {10.1093/mnras/stw1225},
archivePrefix = {arXiv},
       eprint = {1507.01948},
 primaryClass = {astro-ph.GA},
       adsurl = {https://ui.adsabs.harvard.edu/abs/2016MNRAS.460.3100C}
}

@ARTICLE{Zu2022MNRAS,
       author = {{Zu}, Ying and {Song}, Yunjia and {Shao}, Zhiwei and {Chen}, Xiaokai and {Zheng}, Yun and {Gao}, Hongyu and {Yu}, Yu and {Shan}, Huanyuan and {Jing}, Yipeng},
        title = "{Strong conformity and assembly bias: towards a physical understanding of the galaxy-halo connection in SDSS clusters}",
      journal = {Monthly Notices of the Royal Astronomical Society},
         year = 2022,
        month = apr,
       volume = {511},
       number = {2},
        pages = {1789-1807},
          doi = {10.1093/mnras/stac125},
archivePrefix = {arXiv},
       eprint = {2108.06790},
 primaryClass = {astro-ph.CO},
       adsurl = {https://ui.adsabs.harvard.edu/abs/2022MNRAS.511.1789Z}
}

@ARTICLE{zu2020arXiv,
       author = {{Zu}, Ying},
        title = "{On the ``Lensing is Low'' of BOSS Galaxies}",
      journal = {arXiv e-prints},
         year = 2020,
        month = oct,
          eid = {arXiv:2010.01143},
        pages = {arXiv:2010.01143},
          doi = {10.48550/arXiv.2010.01143},
archivePrefix = {arXiv},
       eprint = {2010.01143},
 primaryClass = {astro-ph.CO},
       adsurl = {https://ui.adsabs.harvard.edu/abs/2020arXiv201001143Z}
}

@ARTICLE{Yang2006ApJ,
       author = {{Yang}, Xiaohu and {Mo}, H.~J. and {van den Bosch}, Frank C.},
        title = "{Observational Evidence for an Age Dependence of Halo Bias}",
      journal = {The Astrophysical Journal Letters},
         year = 2006,
        month = feb,
       volume = {638},
       number = {2},
        pages = {L55-L58},
          doi = {10.1086/501069},
archivePrefix = {arXiv},
       eprint = {astro-ph/0509626},
 primaryClass = {astro-ph},
       adsurl = {https://ui.adsabs.harvard.edu/abs/2006ApJ...638L..55Y}
}

@ARTICLE{Sunayama2022arXiv,
       author = {{Sunayama}, Tomomi and {More}, Surhud and {Miyatake}, Hironao},
        title = "{Halo Assembly Bias using properties of central galaxies in SDSS redMaPPer clusters}",
      journal = {arXiv e-prints},
         year = 2022,
        month = may,
          eid = {arXiv:2205.03277},
        pages = {arXiv:2205.03277},
          doi = {10.48550/arXiv.2205.03277},
archivePrefix = {arXiv},
       eprint = {2205.03277},
 primaryClass = {astro-ph.CO},
       adsurl = {https://ui.adsabs.harvard.edu/abs/2022arXiv220503277S}
}

@ARTICLE{Zhang2025Natur,
       author = {{Zhang}, Ziwen and {Chen}, Yangyao and {Rong}, Yu and {Wang}, Huiyuan and {Mo}, Houjun and {Luo}, Xiong and {Li}, Hao},
        title = "{Unexpected clustering pattern in dwarf galaxies challenges formation models}",
      journal = {\nat},
         year = 2025,
        month = jun,
       volume = {642},
       number = {8066},
        pages = {47-52},
          doi = {10.1038/s41586-025-08965-5},
       adsurl = {https://ui.adsabs.harvard.edu/abs/2025Natur.642...47Z}
}

@ARTICLE{Niemiec2018MNRAS,
       author = {{Niemiec}, A. and {Jullo}, E. and {Montero-Dorta}, A.~D. and {Prada}, F. and {Rodriguez-Torres}, S. and {Perez}, E. and {Klypin}, A. and {Erben}, T. and {Makler}, M. and {Moraes}, B. and {Pereira}, M.~E.~S. and {Shan}, H.},
        title = "{Probing galaxy assembly bias with LRG weak lensing observations}",
      journal = {Monthly Notices of the Royal Astronomical Society},
         year = 2018,
        month = jun,
       volume = {477},
       number = {1},
        pages = {L1-L5},
          doi = {10.1093/mnrasl/sly041},
archivePrefix = {arXiv},
       eprint = {1801.06551},
 primaryClass = {astro-ph.CO},
       adsurl = {https://ui.adsabs.harvard.edu/abs/2018MNRAS.477L...1N}
}

@ARTICLE{Wang2022MNRAS,
       author = {{Wang}, Kuan and {Mao}, Yao-Yuan and {Zentner}, Andrew R. and {Guo}, Hong and {Lange}, Johannes U. and {van den Bosch}, Frank C. and {Mezini}, Lorena},
        title = "{Evidence of galaxy assembly bias in SDSS DR7 galaxy samples from count statistics}",
      journal = {Monthly Notices of the Royal Astronomical Society},
         year = 2022,
        month = nov,
       volume = {516},
       number = {3},
        pages = {4003-4024},
          doi = {10.1093/mnras/stac2465},
archivePrefix = {arXiv},
       eprint = {2204.05332},
 primaryClass = {astro-ph.GA},
       adsurl = {https://ui.adsabs.harvard.edu/abs/2022MNRAS.516.4003W}
}

@ARTICLE{Miyatake2016PhRvL,
       author = {{Miyatake}, Hironao and {More}, Surhud and {Takada}, Masahiro and {Spergel}, David N. and {Mandelbaum}, Rachel and {Rykoff}, Eli S. and {Rozo}, Eduardo},
        title = "{Evidence of Halo Assembly Bias in Massive Clusters}",
      journal = {\prl},
         year = 2016,
        month = jan,
       volume = {116},
       number = {4},
          eid = {041301},
        pages = {041301},
          doi = {10.1103/PhysRevLett.116.041301},
archivePrefix = {arXiv},
       eprint = {1506.06135},
 primaryClass = {astro-ph.CO},
       adsurl = {https://ui.adsabs.harvard.edu/abs/2016PhRvL.116d1301M}
}

@ARTICLE{Oyarz2024ApJ,
       author = {{Oyarz{\'u}n}, Grecco A. and {Tinker}, Jeremy L. and {Bundy}, Kevin and {Xhakaj}, Enia and {Wyithe}, J. Stuart B.},
        title = "{Galaxy Assembly Bias in the Stellar-to-halo Mass Relation for Red Central Galaxies from SDSS}",
      journal = {The Astrophysical Journal},
         year = 2024,
        month = oct,
       volume = {974},
       number = {1},
          eid = {29},
        pages = {29},
          doi = {10.3847/1538-4357/ad6de1},
archivePrefix = {arXiv},
       eprint = {2409.03004},
 primaryClass = {astro-ph.GA},
       adsurl = {https://ui.adsabs.harvard.edu/abs/2024ApJ...974...29O}
}

@ARTICLE{Lin2016ApJ,
       author = {{Lin}, Yen-Ting and {Mandelbaum}, Rachel and {Huang}, Yun-Hsin and {Huang}, Hung-Jin and {Dalal}, Neal and {Diemer}, Benedikt and {Jian}, Hung-Yu and {Kravtsov}, Andrey},
        title = "{On Detecting Halo Assembly Bias with Galaxy Populations}",
      journal = {The Astrophysical Journal},
         year = 2016,
        month = mar,
       volume = {819},
       number = {2},
          eid = {119},
        pages = {119},
          doi = {10.3847/0004-637X/819/2/119},
archivePrefix = {arXiv},
       eprint = {1504.07632},
 primaryClass = {astro-ph.GA},
       adsurl = {https://ui.adsabs.harvard.edu/abs/2016ApJ...819..119L}
}

@ARTICLE{Lin2022A&A,
       author = {{Lin}, Yen-Ting and {Miyatake}, Hironao and {Guo}, Hong and {Chiang}, Yi-Kuan and {Chen}, Kai-Feng and {Lan}, Ting-Wen and {Chang}, Yu-Yen},
        title = "{A pair of early- and late-forming galaxy cluster samples: A novel way of studying halo assembly bias assisted by a constrained simulation}",
      journal = {Astronomy \& Astrophysics},
         year = 2022,
        month = oct,
       volume = {666},
          eid = {A97},
        pages = {A97},
          doi = {10.1051/0004-6361/202244404},
archivePrefix = {arXiv},
       eprint = {2202.01795},
 primaryClass = {astro-ph.CO},
       adsurl = {https://ui.adsabs.harvard.edu/abs/2022A&A...666A..97L}
}

@ARTICLE{Lazeyras2017JCAP,
       author = {{Lazeyras}, Titouan and {Musso}, Marcello and {Schmidt}, Fabian},
        title = "{Large-scale assembly bias of dark matter halos}",
      journal = {Journal of Cosmology and Astroparticle Physics},
         year = 2017,
        month = mar,
       volume = {2017},
       number = {3},
          eid = {059},
        pages = {059},
          doi = {10.1088/1475-7516/2017/03/059},
archivePrefix = {arXiv},
       eprint = {1612.04360},
 primaryClass = {astro-ph.CO},
       adsurl = {https://ui.adsabs.harvard.edu/abs/2017JCAP...03..059L}
}

@ARTICLE{Sunayama2016MNRAS,
       author = {{Sunayama}, Tomomi and {Hearin}, Andrew P. and {Padmanabhan}, Nikhil and {Leauthaud}, Alexie},
        title = "{The scale-dependence of halo assembly bias}",
      journal = {Monthly Notices of the Royal Astronomical Society},
         year = 2016,
        month = may,
       volume = {458},
       number = {2},
        pages = {1510-1516},
          doi = {10.1093/mnras/stw332},
archivePrefix = {arXiv},
       eprint = {1509.06417},
 primaryClass = {astro-ph.CO},
       adsurl = {https://ui.adsabs.harvard.edu/abs/2016MNRAS.458.1510S}
}

@ARTICLE{Lim2016MNRAS,
       author = {{Lim}, S.~H. and {Mo}, H.~J. and {Wang}, Huiyuan and {Yang}, Xiaohu},
        title = "{An observational proxy of halo assembly time and its correlation with galaxy properties}",
      journal = {Monthly Notices of the Royal Astronomical Society},
         year = 2016,
        month = jan,
       volume = {455},
       number = {1},
        pages = {499-510},
          doi = {10.1093/mnras/stv2282},
archivePrefix = {arXiv},
       eprint = {1502.01256},
 primaryClass = {astro-ph.GA},
       adsurl = {https://ui.adsabs.harvard.edu/abs/2016MNRAS.455..499L}
}

@ARTICLE{Sunayama2019MNRAS,
       author = {{Sunayama}, Tomomi and {More}, Surhud},
        title = "{On the measurements of assembly bias and splashback radius using optically selected galaxy clusters}",
      journal = {Monthly Notices of the Royal Astronomical Society},
         year = 2019,
        month = dec,
       volume = {490},
       number = {4},
        pages = {4945-4955},
          doi = {10.1093/mnras/stz2832},
archivePrefix = {arXiv},
       eprint = {1905.07557},
 primaryClass = {astro-ph.CO},
       adsurl = {https://ui.adsabs.harvard.edu/abs/2019MNRAS.490.4945S}
}

@ARTICLE{Camargo2025MNRAS,
       author = {{Camargo}, Y.~D. and {Casas-Miranda}, R.~A.},
        title = "{Exploring the link between galaxy assembly and dark matter halo assembly in IllustrisTNG: insights from the Mutual Information}",
      journal = {Monthly Notices of the Royal Astronomical Society},
         year = 2025,
        month = mar,
       volume = {538},
       number = {1},
        pages = {312-326},
          doi = {10.1093/mnras/staf272},
archivePrefix = {arXiv},
       eprint = {2502.06077},
 primaryClass = {astro-ph.GA},
       adsurl = {https://ui.adsabs.harvard.edu/abs/2025MNRAS.538..312C}
}

@ARTICLE{Xu2020MNRAS,
       author = {{Xu}, Xiaoju and {Zheng}, Zheng},
        title = "{Galaxy assembly bias of central galaxies in the Illustris simulation}",
      journal = {Monthly Notices of the Royal Astronomical Society},
         year = 2020,
        month = feb,
       volume = {492},
       number = {2},
        pages = {2739-2754},
          doi = {10.1093/mnras/staa009},
archivePrefix = {arXiv},
       eprint = {1812.11210},
 primaryClass = {astro-ph.GA},
       adsurl = {https://ui.adsabs.harvard.edu/abs/2020MNRAS.492.2739X}
}

@ARTICLE{Wechsler2006ApJ,
       author = {{Wechsler}, Risa H. and {Zentner}, Andrew R. and {Bullock}, James S. and {Kravtsov}, Andrey V. and {Allgood}, Brandon},
        title = "{The Dependence of Halo Clustering on Halo Formation History, Concentration, and Occupation}",
      journal = {The Astrophysical Journal},
         year = 2006,
        month = nov,
       volume = {652},
       number = {1},
        pages = {71-84},
          doi = {10.1086/507120},
archivePrefix = {arXiv},
       eprint = {astro-ph/0512416},
 primaryClass = {astro-ph},
       adsurl = {https://ui.adsabs.harvard.edu/abs/2006ApJ...652...71W}
}

@article{yang2021extended,
  title={An Extended Halo-based Group/Cluster finder: application to the DESI legacy imaging surveys DR8},
  author={Yang, Xiaohu and Xu, Haojie and He, Min and Gu, Yizhou and Katsianis, Antonios and Meng, Jiacheng and Shi, Feng and Zou, Hu and Zhang, Youcai and Liu, Chengze and others},
  journal={The Astrophysical Journal},
  volume={909},
  number={2},
  pages={143},
  year={2021},
  publisher={IOP Publishing}
}

@ARTICLE{yang2006,
       author = {{Yang}, Xiaohu and {Mo}, H.~J. and {van den Bosch}, Frank C. and {Jing}, Y.~P. and {Weinmann}, Simone M. and {Meneghetti}, M.},
        title = "{Weak lensing by galaxies in groups and clusters - I. Theoretical expectations}",
      journal = {Monthly Notices of the Royal Astronomical Society},
         year = 2006,
        month = dec,
       volume = {373},
       number = {3},
        pages = {1159-1172},
          doi = {10.1111/j.1365-2966.2006.11091.x},
archivePrefix = {arXiv},
       eprint = {astro-ph/0607552},
 primaryClass = {astro-ph},
       adsurl = {https://ui.adsabs.harvard.edu/abs/2006MNRAS.373.1159Y}
}

@ARTICLE{mcmc2000,
       author = {{Christensen}, Nelson and {Meyer}, Renate},
        title = "{Bayesian Methods for Cosmological Parameter Estimation from Cosmic Microwave Background Measurements}",
      journal = {arXiv e-prints},
         year = 2000,
        month = jun,
          eid = {astro-ph/0006401},
        pages = {astro-ph/0006401},
          doi = {10.48550/arXiv.astro-ph/0006401},
archivePrefix = {arXiv},
       eprint = {astro-ph/0006401},
 primaryClass = {astro-ph},
       adsurl = {https://ui.adsabs.harvard.edu/abs/2000astro.ph..6401C}
}

@article{van2013,
    author = {van den Bosch, Frank C. and More, Surhud and Cacciato, Marcello and Mo, Houjun and Yang, Xiaohu},
    title = {Cosmological constraints from a combination of galaxy clustering and lensing – I. Theoretical framework},
    journal = {Monthly Notices of the Royal Astronomical Society},
    volume = {430},
    number = {2},
    pages = {725-746},
    year = {2013},
    month = {02},
    issn = {0035-8711},
    doi = {10.1093/mnras/sts006},
    url = {https://doi.org/10.1093/mnras/sts006},
    eprint = {https://academic.oup.com/mnras/article-pdf/430/2/725/9375336/sts006.pdf},
}

@ARTICLE{offcen2007,
       author = {{Johnston}, David E. and {Sheldon}, Erin S. and {Wechsler}, Risa H. and {Rozo}, Eduardo and {Koester}, Benjamin P. and {Frieman}, Joshua A. and {McKay}, Timothy A. and {Evrard}, August E. and {Becker}, Matthew R. and {Annis}, James},
        title = "{Cross-correlation Weak Lensing of SDSS galaxy Clusters II: Cluster Density Profiles and the Mass--Richness Relation}",
      journal = {arXiv e-prints},
         year = 2007,
        month = sep,
          eid = {arXiv:0709.1159},
        pages = {arXiv:0709.1159},
          doi = {10.48550/arXiv.0709.1159},
archivePrefix = {arXiv},
       eprint = {0709.1159},
 primaryClass = {astro-ph},
       adsurl = {https://ui.adsabs.harvard.edu/abs/2007arXiv0709.1159J}
}

@article{navarro1997universal,
  title={A universal density profile from hierarchical clustering},
  author={Navarro, Julio F and Frenk, Carlos S and White, Simon DM},
  journal={The Astrophysical Journal},
  volume={490},
  number={2},
  pages={493},
  year={1997},
  publisher={IOP Publishing}
}

@ARTICLE{Paviot2024A&A,
       author = {{Paviot}, R. and {Rocher}, A. and {Codis}, S. and {de Mattia}, A. and {Jullo}, E. and {de la Torre}, S.},
        title = "{Impact of assembly bias on clustering plus weak lensing cosmological analysis}",
      journal = {Astronomy \& Astrophysics},
         year = 2024,
        month = oct,
       volume = {690},
          eid = {A221},
        pages = {A221},
          doi = {10.1051/0004-6361/202449574},
archivePrefix = {arXiv},
       eprint = {2402.07715},
 primaryClass = {astro-ph.CO},
       adsurl = {https://ui.adsabs.harvard.edu/abs/2024A&A...690A.221P}
}

@ARTICLE{Hearin2016MNRAS,
       author = {{Hearin}, Andrew P. and {Zentner}, Andrew R. and {van den Bosch}, Frank C. and {Campbell}, Duncan and {Tollerud}, Erik},
        title = "{Introducing decorated HODs: modelling assembly bias in the galaxy-halo connection}",
      journal = {Monthly Notices of the Royal Astronomical Society},
         year = 2016,
        month = aug,
       volume = {460},
       number = {3},
        pages = {2552-2570},
          doi = {10.1093/mnras/stw840},
archivePrefix = {arXiv},
       eprint = {1512.03050},
 primaryClass = {astro-ph.CO},
       adsurl = {https://ui.adsabs.harvard.edu/abs/2016MNRAS.460.2552H}
}

@ARTICLE{Bose2019MNRAS,
       author = {{Bose}, Sownak and {Eisenstein}, Daniel J. and {Hernquist}, Lars and {Pillepich}, Annalisa and {Nelson}, Dylan and {Marinacci}, Federico and {Springel}, Volker and {Vogelsberger}, Mark},
        title = "{Revealing the galaxy-halo connection in IllustrisTNG}",
      journal = {Monthly Notices of the Royal Astronomical Society},
         year = 2019,
        month = dec,
       volume = {490},
       number = {4},
        pages = {5693-5711},
          doi = {10.1093/mnras/stz2546},
archivePrefix = {arXiv},
       eprint = {1905.08799},
 primaryClass = {astro-ph.CO},
       adsurl = {https://ui.adsabs.harvard.edu/abs/2019MNRAS.490.5693B}
}

@ARTICLE{Artale2018MNRAS,
       author = {{Artale}, M. Celeste and {Zehavi}, Idit and {Contreras}, Sergio and {Norberg}, Peder},
        title = "{The impact of assembly bias on the halo occupation in hydrodynamical simulations}",
      journal = {Monthly Notices of the Royal Astronomical Society},
         year = 2018,
        month = nov,
       volume = {480},
       number = {3},
        pages = {3978-3992},
          doi = {10.1093/mnras/sty2110},
archivePrefix = {arXiv},
       eprint = {1805.06938},
 primaryClass = {astro-ph.GA},
       adsurl = {https://ui.adsabs.harvard.edu/abs/2018MNRAS.480.3978A}
}

@ARTICLE{LS1993,
       author = {{Landy}, Stephen D. and {Szalay}, Alexander S.},
        title = "{Bias and Variance of Angular Correlation Functions}",
      journal = {The Astrophysical Journal},
         year = 1993,
        month = jul,
       volume = {412},
        pages = {64},
          doi = {10.1086/172900},
       adsurl = {https://ui.adsabs.harvard.edu/abs/1993ApJ...412...64L}
}

@ARTICLE{Wang2022,
       author = {{Wang}, Jiaqi and {Yang}, Xiaohu and {Zhang}, Jun and {Li}, Hekun and {Fong}, Matthew and {Xu}, Haojie and {He}, Min and {Gu}, Yizhou and {Luo}, Wentao and {Dong}, Fuyu and {Wang}, Yirong and {Li}, Qingyang and {Katsianis}, Antonios and {Wang}, Haoran and {Shen}, Zhi and {Alonso Vaquero}, Pedro and {Liu}, Cong and {Huang}, Yiqi and {Liu}, Zhenjie},
        title = "{Halo Properties and Mass Functions of Groups/Clusters from the DESI Legacy Imaging Surveys DR9}",
      journal = {The Astrophysical Journal},
         year = 2022,
        month = sep,
       volume = {936},
       number = {2},
          eid = {161},
        pages = {161},
          doi = {10.3847/1538-4357/ac8986},
archivePrefix = {arXiv},
       eprint = {2207.12771},
 primaryClass = {astro-ph.CO},
       adsurl = {https://ui.adsabs.harvard.edu/abs/2022ApJ...936..161W}
}

@ARTICLE{Springel2018MNRAS,
       author = {{Springel}, Volker and {Pakmor}, R{\"u}diger and {Pillepich}, Annalisa and {Weinberger}, Rainer and {Nelson}, Dylan and {Hernquist}, Lars and {Vogelsberger}, Mark and {Genel}, Shy and {Torrey}, Paul and {Marinacci}, Federico and {Naiman}, Jill},
        title = "{First results from the IllustrisTNG simulations: matter and galaxy clustering}",
      journal = {Monthly Notices of the Royal Astronomical Society},
         year = 2018,
        month = mar,
       volume = {475},
       number = {1},
        pages = {676-698},
          doi = {10.1093/mnras/stx3304},
archivePrefix = {arXiv},
       eprint = {1707.03397},
 primaryClass = {astro-ph.GA},
       adsurl = {https://ui.adsabs.harvard.edu/abs/2018MNRAS.475..676S}
}

@ARTICLE{Planck2016,
       author = {{Planck Collaboration} and {Ade}, P.~A.~R. and {Aghanim}, N. and {Arnaud}, M. and {Ashdown}, M. and {Aumont}, J. and {Baccigalupi}, C. and {Banday}, A.~J. and {Barreiro}, R.~B. and {Bartlett}, J.~G. and {Bartolo}, N. and {Battaner}, E. and {Battye}, R. and {Benabed}, K. and {Beno{\^\i}t}, A. and {Benoit-L{\'e}vy}, A. and {Bernard}, J. -P. and {Bersanelli}, M. and {Bielewicz}, P. and {Bock}, J.~J. and {Bonaldi}, A. and {Bonavera}, L. and {Bond}, J.~R. and {Borrill}, J. and {Bouchet}, F.~R. and {Boulanger}, F. and {Bucher}, M. and {Burigana}, C. and {Butler}, R.~C. and {Calabrese}, E. and {Cardoso}, J. -F. and {Catalano}, A. and {Challinor}, A. and {Chamballu}, A. and {Chary}, R. -R. and {Chiang}, H.~C. and {Chluba}, J. and {Christensen}, P.~R. and {Church}, S. and {Clements}, D.~L. and {Colombi}, S. and {Colombo}, L.~P.~L. and {Combet}, C. and {Coulais}, A. and {Crill}, B.~P. and {Curto}, A. and {Cuttaia}, F. and {Danese}, L. and {Davies}, R.~D. and {Davis}, R.~J. and {de Bernardis}, P. and {de Rosa}, A. and {de Zotti}, G. and {Delabrouille}, J. and {D{\'e}sert}, F. -X. and {Di Valentino}, E. and {Dickinson}, C. and {Diego}, J.~M. and {Dolag}, K. and {Dole}, H. and {Donzelli}, S. and {Dor{\'e}}, O. and {Douspis}, M. and {Ducout}, A. and {Dunkley}, J. and {Dupac}, X. and {Efstathiou}, G. and {Elsner}, F. and {En{\ss}lin}, T.~A. and {Eriksen}, H.~K. and {Farhang}, M. and {Fergusson}, J. and {Finelli}, F. and {Forni}, O. and {Frailis}, M. and {Fraisse}, A.~A. and {Franceschi}, E. and {Frejsel}, A. and {Galeotta}, S. and {Galli}, S. and {Ganga}, K. and {Gauthier}, C. and {Gerbino}, M. and {Ghosh}, T. and {Giard}, M. and {Giraud-H{\'e}raud}, Y. and {Giusarma}, E. and {Gjerl{\o}w}, E. and {Gonz{\'a}lez-Nuevo}, J. and {G{\'o}rski}, K.~M. and {Gratton}, S. and {Gregorio}, A. and {Gruppuso}, A. and {Gudmundsson}, J.~E. and {Hamann}, J. and {Hansen}, F.~K. and {Hanson}, D. and {Harrison}, D.~L. and {Helou}, G. and {Henrot-Versill{\'e}}, S. and {Hern{\'a}ndez-Monteagudo}, C. and {Herranz}, D. and {Hildebrandt}, S.~R. and {Hivon}, E. and {Hobson}, M. and {Holmes}, W.~A. and {Hornstrup}, A. and {Hovest}, W. and {Huang}, Z. and {Huffenberger}, K.~M. and {Hurier}, G. and {Jaffe}, A.~H. and {Jaffe}, T.~R. and {Jones}, W.~C. and {Juvela}, M. and {Keih{\"a}nen}, E. and {Keskitalo}, R. and {Kisner}, T.~S. and {Kneissl}, R. and {Knoche}, J. and {Knox}, L. and {Kunz}, M. and {Kurki-Suonio}, H. and {Lagache}, G. and {L{\"a}hteenm{\"a}ki}, A. and {Lamarre}, J. -M. and {Lasenby}, A. and {Lattanzi}, M. and {Lawrence}, C.~R. and {Leahy}, J.~P. and {Leonardi}, R. and {Lesgourgues}, J. and {Levrier}, F. and {Lewis}, A. and {Liguori}, M. and {Lilje}, P.~B. and {Linden-V{\o}rnle}, M. and {L{\'o}pez-Caniego}, M. and {Lubin}, P.~M. and {Mac{\'\i}as-P{\'e}rez}, J.~F. and {Maggio}, G. and {Maino}, D. and {Mandolesi}, N. and {Mangilli}, A. and {Marchini}, A. and {Maris}, M. and {Martin}, P.~G. and {Martinelli}, M. and {Mart{\'\i}nez-Gonz{\'a}lez}, E. and {Masi}, S. and {Matarrese}, S. and {McGehee}, P. and {Meinhold}, P.~R. and {Melchiorri}, A. and {Melin}, J. -B. and {Mendes}, L. and {Mennella}, A. and {Migliaccio}, M. and {Millea}, M. and {Mitra}, S. and {Miville-Desch{\^e}nes}, M. -A. and {Moneti}, A. and {Montier}, L. and {Morgante}, G. and {Mortlock}, D. and {Moss}, A. and {Munshi}, D. and {Murphy}, J.~A. and {Naselsky}, P. and {Nati}, F. and {Natoli}, P. and {Netterfield}, C.~B. and {N{\o}rgaard-Nielsen}, H.~U. and {Noviello}, F. and {Novikov}, D. and {Novikov}, I. and {Oxborrow}, C.~A. and {Paci}, F. and {Pagano}, L. and {Pajot}, F. and {Paladini}, R. and {Paoletti}, D. and {Partridge}, B. and {Pasian}, F. and {Patanchon}, G. and {Pearson}, T.~J. and {Perdereau}, O. and {Perotto}, L. and {Perrotta}, F. and {Pettorino}, V. and {Piacentini}, F. and {Piat}, M. and {Pierpaoli}, E. and {Pietrobon}, D. and {Plaszczynski}, S. and {Pointecouteau}, E. and {Polenta}, G. and {Popa}, L. and {Pratt}, G.~W. and {Pr{\'e}zeau}, G.},
        title = "{Planck 2015 results. XIII. Cosmological parameters}",
      journal = {Astronomy \& Astrophysics},
         year = 2016,
        month = sep,
       volume = {594},
          eid = {A13},
        pages = {A13},
          doi = {10.1051/0004-6361/201525830},
archivePrefix = {arXiv},
       eprint = {1502.01589},
 primaryClass = {astro-ph.CO},
       adsurl = {https://ui.adsabs.harvard.edu/abs/2016A&A...594A..13P}
}

@ARTICLE{Pakmor2023MNRAS,
       author = {{Pakmor}, R{\"u}diger and {Springel}, Volker and {Coles}, Jonathan P. and {Guillet}, Thomas and {Pfrommer}, Christoph and {Bose}, Sownak and {Barrera}, Monica and {Delgado}, Ana Maria and {Ferlito}, Fulvio and {Frenk}, Carlos and {Hadzhiyska}, Boryana and {Hern{\'a}ndez-Aguayo}, C{\'e}sar and {Hernquist}, Lars and {Kannan}, Rahul and {White}, Simon D.~M.},
        title = "{The MillenniumTNG Project: the hydrodynamical full physics simulation and a first look at its galaxy clusters}",
      journal = {Monthly Notices of the Royal Astronomical Society},
         year = 2023,
        month = sep,
       volume = {524},
       number = {2},
        pages = {2539-2555},
          doi = {10.1093/mnras/stac3620},
archivePrefix = {arXiv},
       eprint = {2210.10060},
 primaryClass = {astro-ph.CO},
       adsurl = {https://ui.adsabs.harvard.edu/abs/2023MNRAS.524.2539P}
}

@ARTICLE{Hern2023MNRAS,
       author = {{Hern{\'a}ndez-Aguayo}, C{\'e}sar and {Springel}, Volker and {Pakmor}, R{\"u}diger and {Barrera}, Monica and {Ferlito}, Fulvio and {White}, Simon D.~M. and {Hernquist}, Lars and {Hadzhiyska}, Boryana and {Delgado}, Ana Maria and {Kannan}, Rahul and {Bose}, Sownak and {Frenk}, Carlos},
        title = "{The MillenniumTNG Project: high-precision predictions for matter clustering and halo statistics}",
      journal = {Monthly Notices of the Royal Astronomical Society},
         year = 2023,
        month = sep,
       volume = {524},
       number = {2},
        pages = {2556-2578},
          doi = {10.1093/mnras/stad1657},
archivePrefix = {arXiv},
       eprint = {2210.10059},
 primaryClass = {astro-ph.CO},
       adsurl = {https://ui.adsabs.harvard.edu/abs/2023MNRAS.524.2556H}
}

@ARTICLE{Dalal2008ApJ,
       author = {{Dalal}, Neal and {White}, Martin and {Bond}, J. Richard and {Shirokov}, Alexander},
        title = "{Halo Assembly Bias in Hierarchical Structure Formation}",
      journal = {The Astrophysical Journal},
         year = 2008,
        month = nov,
       volume = {687},
       number = {1},
        pages = {12-21},
          doi = {10.1086/591512},
archivePrefix = {arXiv},
       eprint = {0803.3453},
 primaryClass = {astro-ph},
       adsurl = {https://ui.adsabs.harvard.edu/abs/2008ApJ...687...12D}
}

@ARTICLE{Luo2018ApJ,
       author = {{Luo}, Wentao and {Yang}, Xiaohu and {Lu}, Tianhuan and {Shi}, Feng and {Zhang}, Jun and {Mo}, H.~J. and {Shu}, Chenggang and {Fu}, Liping and {Radovich}, Mario and {Zhang}, Jiajun and {Li}, Nan and {Sunayama}, Tomomi and {Wang}, Lei},
        title = "{Galaxy-Galaxy Weak-lensing Measurements from SDSS. II. Host Halo Properties of Galaxy Groups}",
      journal = {The Astrophysical Journal},
         year = 2018,
        month = jul,
       volume = {862},
       number = {1},
          eid = {4},
        pages = {4},
          doi = {10.3847/1538-4357/aacaf1},
archivePrefix = {arXiv},
       eprint = {1712.09030},
 primaryClass = {astro-ph.CO},
       adsurl = {https://ui.adsabs.harvard.edu/abs/2018ApJ...862....4L}
}

@article{Lange2017,
    author = {Lange, Johannes U. and van den Bosch, Frank C. and Hearin, Andrew and Campbell, Duncan and Zentner, Andrew R. and Villarreal, Antonia Sierra and Mao, Yao-Yuan},
    title = {Brightest galaxies as halo centre tracers in SDSS DR7},
    journal = {Monthly Notices of the Royal Astronomical Society},
    volume = {473},
    number = {2},
    pages = {2830-2851},
    year = {2017},
    month = {09},
    issn = {0035-8711},
    doi = {10.1093/mnras/stx2434},
    url = {https://doi.org/10.1093/mnras/stx2434},
    eprint = {https://academic.oup.com/mnras/article-pdf/473/2/2830/43989955/mnras\_473\_2\_2830.pdf},
}

@ARTICLE{Hikage2013,
       author = {{Hikage}, Chiaki and {Mandelbaum}, Rachel and {Takada}, Masahiro and {Spergel}, David N.},
        title = "{Where are the Luminous Red Galaxies (LRGs)? Using correlation measurements and lensing to relate LRGs to dark matter haloes}",
      journal = {Monthly Notices of the Royal Astronomical Society},
         year = 2013,
        month = nov,
       volume = {435},
       number = {3},
        pages = {2345-2370},
          doi = {10.1093/mnras/stt1446},
archivePrefix = {arXiv},
       eprint = {1211.1009},
 primaryClass = {astro-ph.CO},
       adsurl = {https://ui.adsabs.harvard.edu/abs/2013MNRAS.435.2345H}
}

@ARTICLE{Wang2014,
       author = {{Wang}, Lei and {Yang}, Xiaohu and {Shen}, Shiyin and {Mo}, H.~J. and {van den Bosch}, Frank C. and {Luo}, Wentao and {Wang}, Yu and {Lau}, Erwin T. and {Wang}, Q.~D. and {Kang}, Xi and {Li}, Ran},
        title = "{Measuring the X-ray luminosities of SDSS DR7 clusters from ROSAT All Sky Survey}",
      journal = {Monthly Notices of the Royal Astronomical Society},
         year = 2014,
        month = mar,
       volume = {439},
       number = {1},
        pages = {611-622},
          doi = {10.1093/mnras/stt2481},
archivePrefix = {arXiv},
       eprint = {1312.7417},
 primaryClass = {astro-ph.CO},
       adsurl = {https://ui.adsabs.harvard.edu/abs/2014MNRAS.439..611W}
}

@ARTICLE{Hoshino2015,
       author = {{Hoshino}, Hanako and {Leauthaud}, Alexie and {Lackner}, Claire and {Hikage}, Chiaki and {Rozo}, Eduardo and {Rykoff}, Eli and {Mandelbaum}, Rachel and {More}, Surhud and {More}, Anupreeta and {Saito}, Shun and {Vulcani}, Benedetta},
        title = "{Luminous red galaxies in clusters: central occupation, spatial distributions and miscentring}",
      journal = {Monthly Notices of the Royal Astronomical Society},
         year = 2015,
        month = sep,
       volume = {452},
       number = {1},
        pages = {998-1013},
          doi = {10.1093/mnras/stv1271},
archivePrefix = {arXiv},
       eprint = {1503.05200},
 primaryClass = {astro-ph.CO},
       adsurl = {https://ui.adsabs.harvard.edu/abs/2015MNRAS.452..998H}
}

@article{diemer2018colossus,
  title={COLOSSUS: A python toolkit for cosmology, large-scale structure, and dark matter halos},
  author={Diemer, Benedikt},
  journal={The Astrophysical Journal Supplement Series},
  volume={239},
  number={2},
  pages={35},
  year={2018},
  publisher={IOP Publishing}
}

@article{tinker2010large,
  title={The large-scale bias of dark matter halos: numerical calibration and model tests},
  author={Tinker, Jeremy L and Robertson, Brant E and Kravtsov, Andrey V and Klypin, Anatoly and Warren, Michael S and Yepes, Gustavo and Gottl{\"o}ber, Stefan},
  journal={The Astrophysical Journal},
  volume={724},
  number={2},
  pages={878},
  year={2010},
  publisher={IOP Publishing}
}

@article{Zhang2017,
doi = {10.3847/1538-4357/834/1/8},
url = {https://dx.doi.org/10.3847/1538-4357/834/1/8},
year = {2016},
month = {dec},
publisher = {The American Astronomical Society},
volume = {834},
number = {1},
pages = {8},
author = {Jun Zhang and Pengjie Zhang and Wentao Luo},
title = {APPROACHING THE CRAMÉR–RAO BOUND IN WEAK LENSING WITH PDF SYMMETRIZATION},
journal = {The Astrophysical Journal}
}

@article{Li_2021,
doi = {10.3847/1538-4357/abcda3},
url = {https://dx.doi.org/10.3847/1538-4357/abcda3},
year = {2021},
month = {feb},
publisher = {The American Astronomical Society},
volume = {908},
number = {1},
pages = {93},
author = {Hekun Li and Jun Zhang and Dezi Liu and Wentao Luo and Jiajun Zhang and Fuyu Dong and Zhi Shen and Haoran Wang},
title = {Toward a Bias-free Selection Criterion in Shear Measurement},
journal = {The Astrophysical Journal}
}

@ARTICLE{Rongpu2021MNRAS,
       author = {{Zhou}, Rongpu and {Newman}, Jeffrey A. and {Mao}, Yao-Yuan and {Meisner}, Aaron and {Moustakas}, John and {Myers}, Adam D. and {Prakash}, Abhishek and {Zentner}, Andrew R. and {Brooks}, David and {Duan}, Yutong and {Landriau}, Martin and {Levi}, Michael E. and {Prada}, Francisco and {Tarle}, Gregory},
        title = "{The clustering of DESI-like luminous red galaxies using photometric redshifts}",
      journal = {Monthly Notices of the Royal Astronomical Society},
         year = 2021,
        month = mar,
       volume = {501},
       number = {3},
        pages = {3309-3331},
          doi = {10.1093/mnras/staa3764},
archivePrefix = {arXiv},
       eprint = {2001.06018},
 primaryClass = {astro-ph.CO},
       adsurl = {https://ui.adsabs.harvard.edu/abs/2021MNRAS.501.3309Z}
}

@article{Zhang2022,
doi = {10.3847/1538-3881/ac84d8},
url = {https://dx.doi.org/10.3847/1538-3881/ac84d8},
year = {2022},
month = {sep},
publisher = {The American Astronomical Society},
volume = {164},
number = {4},
pages = {128},
author = {Jun Zhang and Cong Liu and Pedro Alonso Vaquero and Hekun Li and Haoran Wang and Zhi Shen and Fuyu Dong},
title = {Shear Measurement with Poorly Resolved Images},
journal = {The Astronomical Journal}
}

@ARTICLE{Zhang2015JCAP,
       author = {{Zhang}, Jun and {Luo}, Wentao and {Foucaud}, Sebastien},
        title = "{Accurate shear measurement with faint sources}",
      journal = {Journal of Cosmology and Astroparticle Physics},
         year = 2015,
        month = jan,
       volume = {2015},
       number = {1},
        pages = {024-024},
          doi = {10.1088/1475-7516/2015/01/024},
archivePrefix = {arXiv},
       eprint = {1312.5514},
 primaryClass = {astro-ph.CO},
       adsurl = {https://ui.adsabs.harvard.edu/abs/2015JCAP...01..024Z}
}

@article{scikit-learn,
 title={Scikit-learn: Machine Learning in {P}ython},
 author={Pedregosa, F. and Varoquaux, G. and Gramfort, A. and Michel, V.
         and Thirion, B. and Grisel, O. and Blondel, M. and Prettenhofer, P.
         and Weiss, R. and Dubourg, V. and Vanderplas, J. and Passos, A. and
         Cournapeau, D. and Brucher, M. and Perrot, M. and Duchesnay, E.},
 journal={the Journal of machine Learning research},
 volume={12},
 pages={2825--2830},
 year={2011}
}

@ARTICLE{Sheth2004MNRAS,
       author = {{Sheth}, Ravi K. and {Tormen}, Giuseppe},
        title = "{On the environmental dependence of halo formation}",
      journal = {Monthly Notices of the Royal Astronomical Society},
         year = 2004,
        month = jun,
       volume = {350},
       number = {4},
        pages = {1385-1390},
          doi = {10.1111/j.1365-2966.2004.07733.x},
archivePrefix = {arXiv},
       eprint = {astro-ph/0402237},
 primaryClass = {astro-ph},
       adsurl = {https://ui.adsabs.harvard.edu/abs/2004MNRAS.350.1385S}
}

\end{document}